\documentclass[a4paper,onecolumn]{aa} 

\usepackage{natbib}
\usepackage{graphicx}
\usepackage{txfonts}

\newcommand{\half}{\ensuremath{\frac{1}{2}}}

\newcommand{\etal}{\textit{et al.\,}}

\newcommand{\vecbf}[1]{\ensuremath{\mathbf{\stackrel{\rightarrow}{#1}}}}

\newcounter{cureqno}%
\newenvironment{mathletters}{%
 \refstepcounter{equation}%
 \setcounter{cureqno}{\value{equation}}%
 \edef\@tempa{\theequation}%
 \expandafter\def
 \expandafter\theequation
 \expandafter{\@tempa\alph{equation}}%
 \setcounter{equation}{0}%
}{%
 \setcounter{equation}{\value{cureqno}}%
}%

\bibpunct[; ]{(}{)}{;}{a}{}{,}

\begin{document}

\title{Photon orbital angular momentum and torque metrics for single telescopes
and interferometers}

\author{N.M. Elias II}

\institute{National Radio Astronomy Observatory\footnote{The National Radio
Astronomy Observatory is a facility of the National Science Foundation operated
under cooperative agreement by Associated Universities, Inc.}; P.V. Domenici
Science Operations Center; P.O. Box O; 1003 Lopezville Road; Socorro, NM
87801-0387; USA\\
\email{nelias@nrao.edu}}

\date{Received 2012 January 03; Accepted 2012 March 08}

\abstract
{Photon orbital angular momentum (POAM) is normally invoked in a quantum
mechanical context.  It can, however, also be adapted to the classical regime,
which includes observational astronomy.}
{I explain why POAM quantities are excellent metrics for describing the
end-to-end behavior of astronomical systems.  To demonstrate their utility, I
calculate POAM probabilities and torques from holography measurements of EVLA
antenna surfaces.}
{With previously defined concepts and calculi, I present generic expressions for
POAM spectra, total POAM, torque spectra, and total torque in the image plane.
I extend these functional forms to describe the specific POAM behavior of both
single telescopes and interferometers.}
{POAM probabilities of spatially uncorrelated astronomical sources are symmetric
in quantum number.  Such objects thus have zero intrinsic total POAM on the
celestial sphere, which means that the total POAM in the image plane is
identical to the total torque induced by aberrations within propagation media
and instrumentation.  The total torque can be divided into source- independent
and dependent components, and the latter can be written in terms of three
illustrative forms.  For interferometers, complications arise from discrete
sampling of synthesized apertures, but they can be overcome.  POAM also
manifests itself in the apodization of each telescope in an array.  Holography
measurements of EVLA antennas observing a point source indicate that $\sim$ 10\%
of photons in the $n$ $=$ $0$ state are torqued to $n$ $\neq$ $0$ states.}
{POAM quantities represent excellent metrics for characterizing instruments
because they are based on real physics and are used to simultaneously describe
amplitude and phase aberrations.  In contrast, Zernike polynomials are just
solutions of a differential equation that happen to $\sim$ correspond to
specific types of aberrations (e.g., tip-tilt, focus, etc.) and are typically
employed to fit only phases.  Possible future studies include forming POAM
quantities with real interferometry visibility data, modeling instrumental
aberrations and turbulence of the troposphere/ionosphere in terms of POAM,
POAM-based imaging algorithms and constraints, POAM-based super resolution
imaging, and POAM observations of astrophysically important sources.} 


\maketitle

\section{Introduction} \label{Sec:Introduction}

\citet{Elias08} developed extensive semi-classical formalisms to describe photon
orbital angular momentum (POAM) in astronomy.  He assumed spatially incoherent
sources, so these formalisms are significantly different from those used in
laboratory situations.  He concentrated more on instrumentation rather than
astrophysics, and included first principles, concepts, definitions, calculi,
examples, and applications.

As a general rule, aberrations that damage wavefronts and POAM spectra should be
minimized in both hardware and software: \textit{Primum non
torquere\footnote{This Latin phrase -- loosely translated as ``Above all, apply
no torque'' -- is a shameless adaptation of the most widely quoted words from
the Hippocratic Oath ``Primum non nocere,'' which means ``Above all, do no
harm.''}}.  The amount of damage to wavefronts and POAM spectra should be
estimated and removed by off-line image processing algorithms whenever possible.
I now extend previous POAM work toward imaging metrics for single telescopes and
interferometers, leading to a deeper understanding of how such instruments work.

\section{Basic Concepts} \label{Sec:BasicConcepts}

The apertures of all real astronomical telescopes are finite in size, so they
are only capable of producing diffraction-limited images up to a resolution of
$\sim$ $D^{-1}$ ($D$ is the aperture diameter expressed in wavelengths).
Assuming that the aperture response is uniform or at least azimuthally symmetric
(radially apodized), this loss of information manifests itself as blurring.  The
only rigorous way to reduce blurring is to increase the aperture size.

Conversely, if the aperture response is azimuthally asymmetric (or, azimuthally
apodized), the image is subject to non-uniform distortion.  These distortions
may occur over all angular scales larger than $\sim$ $D^{-1}$.  The only
rigorous way to reduce distortion is to eliminate the asymmetry in the aperture
response.

I distinguish between these two types of image degradation because it is much
harder to build a very large telescope compared to figuring a very smooth mirror
of nominal size.  There is yet another reason for making this distinction,
namely that radial apodization \textit{does not} modulate POAM spectra, while
azimuthal apodization \textit{does} modulate POAM spectra.  In other words,
aberrations modulate each input POAM state into one or more output POAM states
\citep{Elias08} and introduce image distortion.

Interferometers measure visibilities at discrete points in the synthesized
aperture.  In turn, these measurements are transformed into ``dirty'' images of
astronomical sources.  This process is mathematically equivalent to punching
small pinholes into the opaque aperture of a large single telescope and imaging
the resulting interference optomechanically.

The act of aperture sampling is a form of azimuthal apodization that modifies
POAM spectra.  Other physical effects -- including instrumental imperfections,
turbulence in the troposphere or ionosphere, etc. -- manifest themselves as
sample amplitude and phase errors and also modify POAM quantities. Symmetrically
reducing the weighting of long baselines, on the other hand, only blurs an image
and doesn't modulate POAM quantities.

I can press the analogies even further.  Image-processing algorithms, such as
CLEAN and MEM, remove artifacts to yield a model of the true source up to a
certain resolution.  These artifacts are produced by asymmetric apodization,
which means that image-processing algorithms actually estimate and eliminate
changes in POAM spectra.

\section{Definitions} \label{Sec:Definitions}

According to \citet{Elias08}, the total POAM of a wavefront arising from the
celestial sphere, in units of $\hbar$, is
\begin{mathletters} \label{Ml:POAMCS}
\begin{equation} \label{Eq:POAMCS}
l_{Z} \, = \, L_{Z} / \hbar \, = \, \sum_{m=-\infty}^{\infty} m \, p_{m,m} \, ,
\end{equation}
where
\begin{equation} \label{Eq:pMM}
p_{m,m} \, = \,
B_{m,m} \, / B \, = \,
B_{m,m} \, / \sum_{m=-\infty}^{\infty} B_{m,m}
\end{equation}
is the probability that a single photon (or the fraction of many photons) is in
POAM state $m$, $B_{m,m}$ is the $(m,m)^{th}$ POAM autocorrelation, and $B$ is
the total intensity.  The ensemble of probabilities represents the intrinsic
source POAM spectrum.  The total intensity, which is integrated over the
celestial sphere, is identical to the sum over all POAM autocorrelations.  The
same formulae can be used to describe the total POAM of a wavefront incident
upon the image plane by replacing $l_{Z}$ $\rightarrow$ $\tilde{l}_{Z}$, $L_{Z}$
$\rightarrow$ $\tilde{L}_{Z}$, $p_{m,m}$ $\rightarrow$ $\tilde{p}_{m,m}$,
$B_{m,m}$ $\rightarrow$ $\tilde{B}_{m,m}$, and $B$ $\rightarrow$ $\tilde{B}$.
\end{mathletters}

I define the total torque as the difference of the total POAM between the image
plane and celestial sphere, or
\begin{equation} \label{Eq:TelTorque}
\tau \, = \,
\tilde{l}_{Z} \, - \, l_{Z} \, = \,
\sum_{m=-\infty}^{\infty} m \, \tau_{m,m} \, = \,
\sum_{m=-\infty}^{\infty} m \, \left(
  \tilde{p}_{m,m} \, - \, p_{m,m}
\right) \, ,
\end{equation}
where the ensemble of $\tau_{m,m}$ comprise the torque spectrum.  Torque is
normally defined as change in angular momentum per unit time.  Since ``per unit
time'' is ambiguous in this context I ignore it, so torque has the same units as
POAM.

The electric fields of a ``natural light'' astronomical source projected onto
the celestial sphere are spatially uncorrelated.  I call this scenario the
``Standard Astronomical Assumption'' (SAA), which led to the creation of the
POAM calculi \citep[Tables 1-4]{Elias08}.  Because of SAA, $B_{-m,-m}$ $=$
$B_{m,m}$ and $p_{-m,-m}$ $=$ $p_{m,m}$ for all $m$, implying that $l_{Z}$ $=$
$0$.  I derive this result in Appendix \ref{App:POAMCS}.  Equation
\ref{Eq:TelTorque} then becomes
\begin{equation} \label{Eq:POAMTorque}
\tau \, = \,
\tilde{l}_{Z} \, = \,
\sum_{m=-\infty}^{\infty} m \, \tau_{m,m} \, = \,
\sum_{m=-\infty}^{\infty} m \, \tilde{p}_{m,m} \, .
\end{equation}
Torque applied by propagation media and instrumentation modifies the source POAM
spectral components as they travel to the image plane, but the total POAM in the
image plane does not depend on the total POAM from the celestial sphere because
of the symmetry in $m$.  Although this equation is independent of $l_{Z}$, the
$\pm$ $1$ transitional probabilities of the source do affect $\tilde{l}_{Z}$
(cf. Section \ref{Sec:Telescope}) except for point sources at the center of the
field of view (FOV).

Maser photons traveling through turbulent gas and photons scattering off Kerr
black holes \citep{Harwit03,TTM-TA} may not satisfy the SAA condition.
Therefore, these sources can exhibit $l_{Z}$ $\neq$ $0$ and are beyond the scope
of this paper.

\section{Single Telescopes} \label{Sec:Telescope}

I derive POAM quantities for single telescopes first because it is relatively
simple to do and the results can be extended to other types of instruments
(e.g., interferometers; cf. Section \ref{Sec:Interferometer}).  \citet{Elias08}
created generic calculi that can describe the POAM response of any instrument,
so I will use them here.

\subsection{Initial Mathmematics} \label{SSec:MathTel}

\begin{figure}[!ht]
  \begin{center}
    \includegraphics[width=18cm,angle=0]{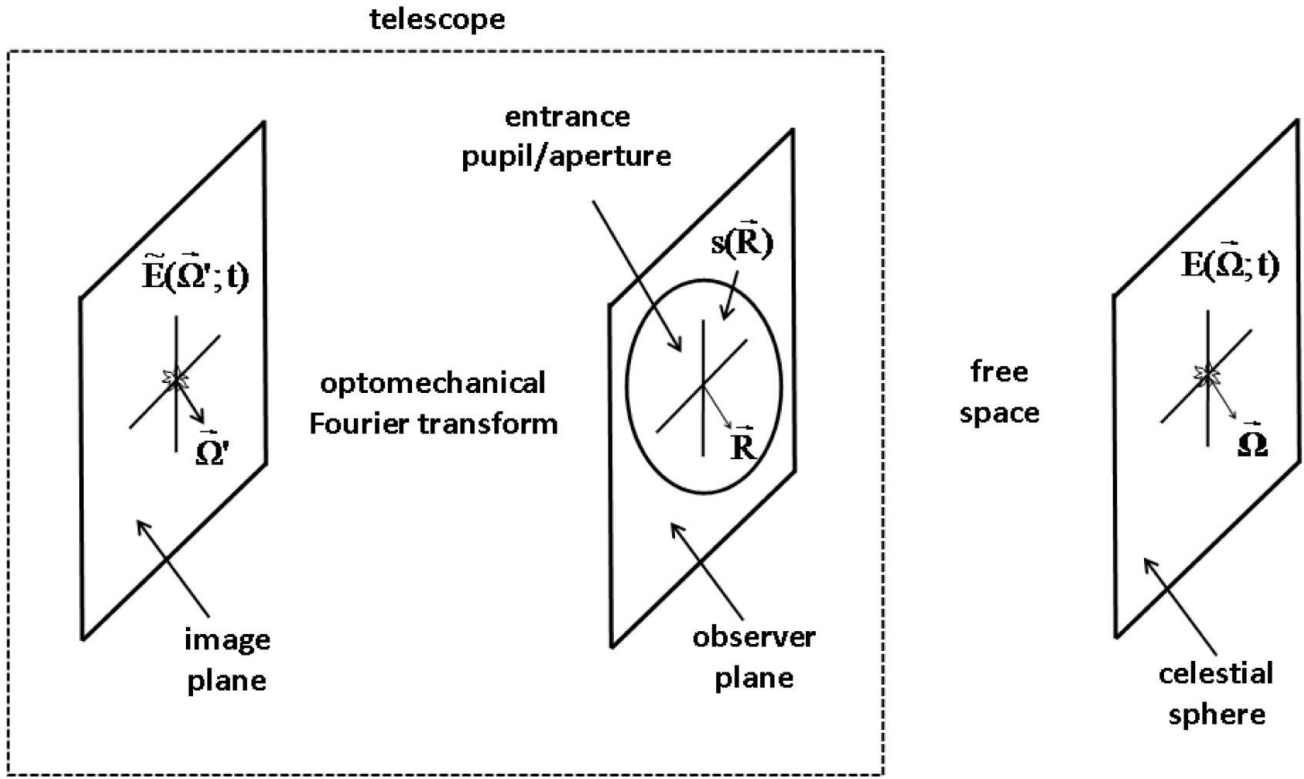}
  \end{center}
  \caption{Schematic diagram of a single telescope looking at a source on the
celestial sphere.  The coordinates are indicated and described in the text of
Section \ref{Sec:Telescope}.  The electric fields (celestial sphere and image
plane) and aperture function are also presented.}
  \label{Fig:Telescope}
\end{figure}

Consider Figure \ref{Fig:Telescope}, the schematic diagram of a single telescope
looking at an object on the celestial sphere.  The intensity response, in system
form, for an SAA source is given by
\begin{mathletters}
\begin{eqnarray} \label{Eq:TelResponse}
\tilde{B}(\vecbf{\Omega}^{\prime}) &=
&\left<
  \frac{1}{2} \left| \tilde{E}(\vecbf{\Omega}^{\prime};t) \right|^{2}
\right> \, = \,
  \left< \frac{1}{2}
    \left|
      \int d^{2}\Omega \, D(\vecbf{\Omega}^{\prime},\vecbf{\Omega}) \,
      E(\vecbf{\Omega};t)
    \right|^{2}
  \right> \nonumber \\ &=
&\int d^{2}\Omega \, \left|
  D(\vecbf{\Omega}^{\prime},\vecbf{\Omega})
\right|^{2} \, \left<
  \frac{1}{2} \left| E(\vecbf{\Omega};t) \right|^{2}
\right> \, = \, \int d^{2}\Omega \, P(\vecbf{\Omega}^{\prime},\vecbf{\Omega}) \,
  B(\vecbf{\Omega}) \, ,
\end{eqnarray}
where $\vecbf{\Omega}$ $=$ $(\rho \cos{\phi}, \rho \sin{\phi})$ is the
coordinate on the celestial sphere, $\vecbf{\Omega}^{\prime}$ $=$
$(\rho^{\prime} \cos{\phi^{\prime}}, \rho^{\prime} \sin{\phi^{\prime}})$ is the
coordinate in the image plane, $\tilde{B}(\vecbf{\Omega}^{\prime})$ and
$B(\vecbf{\Omega})$ are the intensity distributions
($\tilde{E}(\vecbf{\Omega}^{\prime};t)$ and $E(\vecbf{\Omega};t)$ are the
corresponding electric fields), $P(\vecbf{\Omega}^{\prime},\vecbf{\Omega})$ $=$
$P(\vecbf{\Omega}^{\prime}-\vecbf{\Omega})$ is the point-spread function (PSF),
\begin{equation} \label{Eq:DF}
D(\vecbf{\Omega}^{\prime},\vecbf{\Omega}) \, = \,
  D(\vecbf{\Omega}^{\prime}-\vecbf{\Omega}) \, = \,
  \int d^{2}R \,
  e^{-j 2\pi \vecbf{R} \cdot (\vecbf{\Omega}^{\prime} - \vecbf{\Omega})} \,
  s(\vecbf{R})
\end{equation}
is the diffraction functon (DF), $\vecbf{R}$ $=$ $(R\cos{\psi},R\sin{\psi})$ is
the coordinate in the aperture (in units of wavelength), and $s(\vecbf{R})$ is
the functional description of the aperture apodization (cf. Section
\ref{Sec:BasicConcepts}).  Any optical system, including propagation media and
instrumentation, that can be expressed in this mathematical form can be expanded
into any of the POAM calculi.
\end{mathletters}

From \citet[Row 3 of Table 3]{Elias08}, the $(m,m)^{th}$ single-telescope SAA
POAM autocorrelation density is
\begin{mathletters} \label{Ml:BTildeMM}
\begin{eqnarray} \label{Eq:BTildeMM}
\tilde{B}_{m,m}(\rho^{\prime}) \, = \,
  \left<
    \frac{1}{2} \left| \tilde{E}_{m}(\rho^{\prime};t) \right|^{2}
  \right> \, = \,
\int d^{2}\Omega \, P_{m,m}(\rho^{\prime},\vecbf{\Omega}) \,
  B(\vecbf{\Omega}) \, ,
\end{eqnarray}
where
\begin{equation} \label{Eq:ETildeM}
\tilde{E}_{m}(\rho^{\prime};t) \, = \,
  \frac{1}{2\pi} \int_{0}^{2\pi} d\phi^{\prime} \, e^{-jm \phi^{\prime}} \,
  \tilde{E}(\vecbf{\Omega}^{\prime};t) \, = \,
\int d^{2}\Omega \, D_{m}(\rho^{\prime},\vecbf{\Omega}) \, E(\vecbf{\Omega};t)
  ~~ \stackrel{\mathcal{F}}{\Leftrightarrow} ~~
\tilde{E}(\vecbf{\Omega}^{\prime};t) \, = \,
  \sum_{m=-\infty}^{\infty} \tilde{E}_{m}(\rho^{\prime};t) \,
  e^{jm \phi^{\prime}}
\end{equation}
is the $m^{th}$ image-plane POAM state,
\begin{equation} \label{Eq:TelPMM}
P_{m,m}(\rho^{\prime},\vecbf{\Omega}) \, = \,
  \left| D_{m}(\rho^{\prime},\vecbf{\Omega}) \right|^{2} \, = \,
\left|
  \sum_{k=-\infty}^{\infty} j^{k} \, \mathcal{J}_{m,k}(\rho^{\prime},\rho) \,
  e^{-jk \phi}
\right|^{2}
\end{equation}
is the $(m,m)^{th}$ PSF sensitivity,
\begin{equation} \label{Eq:TelDM}
D_{m}(\rho^{\prime},\vecbf{\Omega}) \, = \,
  \frac{1}{2\pi} \int_{0}^{2\pi} d\phi^{\prime} \, e^{-jm \phi^{\prime}} \,
  D(\vecbf{\Omega}^{\prime},\vecbf{\Omega})
  ~~ \stackrel{\mathcal{F}}{\Leftrightarrow} ~~
D(\vecbf{\Omega}^{\prime},\vecbf{\Omega}) \, = \,
  \sum_{m=-\infty}^{\infty} D_{m}(\rho^{\prime},\vecbf{\Omega}) \,
  e^{jm \phi^{\prime}}
\end{equation}
is the $m^{th}$ DF sensitivity,
\begin{equation} \label{Eq:TelIntFunc}
\mathcal{J}_{p,q}(\rho^{\prime},\rho) \, = \,
  2\pi \int_{0}^{R_{tel}} dR \, R \, J_{p}(2\pi R \rho^{\prime}) \, s_{p-q}(R)
  \, J_{q}(2\pi R \rho)
\end{equation}
is the $(p,q)^{th}$ integral function, $R_{tel}$ is the telescope radius,
$J_{g}(x)$ is the $g^{th}$ order Bessel function of the first kind, and
\begin{equation} \label{Eq:TelApFuncG}
s_{g}(R) \, = \,
  \frac{1}{2\pi} \int_{0}^{2\pi} d\psi \, e^{-j g \psi} \, s(\vecbf{R})
  ~ \stackrel{\mathcal{F}}{\Leftrightarrow} ~
s(\vecbf{R}) \, = \,
  \sum_{g=-\infty}^{\infty} s_{g}(R) \, e^{j g \psi}
\end{equation}
is the $g^{th}$ azimuthal Fourier component of the aperture function
$s(\vecbf{R})$.  Integrating Equation \ref{Eq:BTildeMM} over radius in the image
plane leads to the $(m,m)^{th}$ POAM autocorrelation
\end{mathletters}
\begin{mathletters} \label{Ml:BTildeMMTotal}
\begin{eqnarray} \label{Eq:BTildeMMTotal}
\tilde{B}_{m,m} \, = \,
  \lim_{\rho_{FOV} \rightarrow \infty} 2\pi \int_{0}^{\rho_{FOV}} d\rho^{\prime}
  \, \rho^{\prime} \, \tilde{B}_{m,m}(\rho^{\prime}) \, = \,
\int d^{2}\Omega \, P_{m,m}(\vecbf{\Omega}) \, B(\vecbf{\Omega}) \, ,
\end{eqnarray}
where $\rho_{FOV}$ is the FOV of the image plane, and
\begin{equation} \label{Eq:TelPMMTotal}
P_{m,m}(\vecbf{\Omega}) \, = \,
\lim_{\rho_{FOV} \rightarrow \infty} 2\pi \int_{0}^{\rho_{FOV}} d\rho^{\prime}
  \, \rho^{\prime} \, P_{m,m}(\rho^{\prime},\vecbf{\Omega}) \, = \,
\sum_{k=-\infty}^{\infty} \sum_{l=-\infty}^{\infty} j^{k-l} \, 2\pi
  \int_{0}^{R_{tel}} dR \, R \, s_{m-k}(R) \, s^{*}_{m-l}(R) \,
  J_{k}(2\pi R \rho) \, J_{l}(2\pi R \rho) \, e^{-j(k-l)\phi} ~~~~~~
\end{equation}
is the $(m,m)^{th}$ PSF sensitivity kernel.  I assume that $\rho_{FOV}$ is large
enough to capture most of the radiation scattered through the telescope aperture
into the image plane.  The complete derivation of Equations
\ref{Ml:BTildeMM}-\ref{Ml:BTildeMMTotal} may be found in Appendix
\ref{App:TelCorr}.
\end{mathletters}

All POAM quantities defined in Section \ref{Sec:Definitions} can be formed from
Equations \ref{Ml:BTildeMMTotal}a-b.  For example, the total POAM in the image
plane is
\begin{eqnarray} \label{Eq:TildeLz}
\tilde{l}_{Z} \, = \,
  \frac{1}{\tilde{B}} \sum_{m=-\infty}^{\infty} m \, \tilde{B}_{m,m} \, = \,
  \frac{1}{\tilde{B}} \int d^{2}\Omega \,
    \left[
      \sum_{m=-\infty}^{\infty} m \, P_{m,m}(\vecbf{\Omega})
    \right] \, B(\vecbf{\Omega}) \, = \,
  \frac{1}{\tilde{B}} \int d^{2}\Omega \,
    \tilde{\mathcal{L}}_{z}(\vecbf{\Omega}) \, B(\vecbf{\Omega}) \, ,
\end{eqnarray}
where $\tilde{\mathcal{L}}_{Z}(\vecbf{\Omega})$ is the total POAM kernel.  Total
POAM on the celestial sphere for all SAA sources is identically zero, or $l_{Z}$
$=$ $0$ (cf. Appendix \ref{App:POAMCS}).  This statement means that the total
POAM measured in the image plane is identical to total torque applied to the
wavefronts.  Therefore, I interchangably employ the quantities $\tilde{l}_{Z}$
$\leftrightarrow$ $\tau$ and $\tilde{\mathcal{L}}_{Z}(\vecbf{\Omega})$
$\leftrightarrow$ $\mathcal{T}(\vecbf{\Omega})$, where
$\mathcal{T}(\vecbf{\Omega})$ is the total torque kernel.  For a more physical
understanding of Equation \ref{Eq:TildeLz}, Equations \ref{Ml:BTildeMMTotal}a-b
are combined with various mathematical identities to create three ``illustrative
forms'' of $\tilde{l}_{Z}$ which emphasize different aspects of POAM for single
telescopes (cf. Section \ref{SSec:IllForms}).

\subsection{Illustrative Forms of $\tilde{l}_{Z}$} \label{SSec:IllForms}

The first illustrative form of $\tilde{l}_{Z}$ is
\begin{eqnarray} \label{Eq:TelTorque1}
\tilde{l}_{Z} =
\sum_{m=-\infty}^{\infty} m p^{a}_{m,m} \pm
\mathrm{Im} \, 2\pi \int_{0}^{\infty} d\rho \, \rho \, 2\pi
  \int_{0}^{R_{tel}} dR \, R \, \left( 2\pi R \rho \right)
  \left[ \sum_{n=-\infty}^{\infty} p_{n,n \pm 1}(\rho) \right]
  \left[ \sum_{m=-\infty}^{\infty} p^{a}_{m,m \mp 1}(R) \right] ,
\end{eqnarray}
where $p^{a}_{m,m}$ is the $m^{th}$ POAM state probability in the aperture,
$p_{n,n \pm 1}(\rho)$ is the transitional probability density between POAM
states $n$ and $n$ $\pm$ 1 on the celestial sphere, and $p^{a}_{m,m \mp 1}(R)$
is the transitional probability density between POAM states $m$ and $m$ $\mp$
$1$ in the aperture.  The transitional probabilities correspond to $\pm 1$
selection rules.  I derive these equations and the variables contained therein
in Appendix
\ref{App:IllForm}.

The second illustrative form of $\tilde{l}_{Z}$ is
\begin{mathletters}
\begin{equation} \label{Eq:TelTorque2}
\tilde{l}_{Z} \, = \,
  \sum_{m=-\infty}^{\infty} m \, p^{a}_{m,m} \, \pm \, \mathrm{Im} \,
  2\pi \int_{0}^{\infty} d\rho \, \rho ~ 2\pi \int_{0}^{R_{tel}} \, dR \, R \,
  \left( 2\pi R \rho \right) \,
  \left[
   \frac{\mathcal{B}_{\mp 1}(\rho)}{B}
  \right]
  \left[
    \frac{\mathcal{S}_{\pm 1}(R)}{S}
  \right] \, ,
\end{equation}
where the $\mathcal{B}_{\mp 1}(\rho)$ are the first order rancors of the source,
the $\mathcal{S}_{\pm 1}(R)$ are the first order rancor sensitivities, and $S$
is the integrated squared magnitude of the aperture function.  This expression
proves that rancors and rancor sensitivities \citep{Elias08},
\begin{equation} \label{Eq:R2}
\mathcal{B}_{p}(\rho) =
  \frac{1}{2\pi} \int_{0}^{2\pi} d\phi \, e^{-jp\phi} \, B(\vecbf{\Omega}) =
  \sum_{m=-\infty}^{\infty} B_{m,m-p}(\rho)
~~\, \stackrel{\mathcal{F}}{\Leftrightarrow} ~~
B(\vecbf{\Omega}) \, = \, 
  \sum_{p=-\infty}^{\infty} \mathcal{B}_{p}(\rho) \, e^{jp\phi}
~~~~ \left[
  B_{i,k}(\rho) \, = \,
    \left< \frac{1}{2} E_{i}(\rho;t) E^{*}_{k}(\rho;t) \right>
  \right] , ~
\end{equation}
\begin{equation} \label{Eq:R2}
\mathcal{S}_{p}(R) =
  \frac{1}{2\pi} \int_{0}^{2\pi} d\psi \, e^{-jp\psi} \, S(\vecbf{R}) =
  \sum_{m=-\infty}^{\infty} S_{m,m-p}(R)
~~\, \stackrel{\mathcal{F}}{\Leftrightarrow} ~~
S(\vecbf{R}) \, = \, 
  \sum_{p=-\infty}^{\infty} \mathcal{S}_{p}(R) \, e^{jp\psi}
~~~~ \left[
    S_{i,k}(R) \, = \, s_{i}(R) s^{*}_{k}(R) , ~
    S(\vecbf{R}) \, = \, \left|s(\vecbf{R})\right|^2
  \right] , ~
\end{equation}
\end{mathletters}
calculated directly from intensities and squared aperture functions (instead of
electric fields and aperture functions) and related to the transitional
probabilities of Equation \ref{Eq:TelTorque1}, are relevant for POAM analysis.
I derive these equations and the variables contained therein in Appendix
\ref{App:IllForm}.

The third illustrative form of $\tilde{l}_{Z}$ is
\begin{equation} \label{Eq:TelTorque3}
\tilde{l}_{Z} \, = \,
\sum_{m=-\infty}^{\infty} m \, p^{a}_{m,m} \, + \,
  2\pi \int d^{2}\Omega \int d^{2}R \,
  \left[ \vecbf{\Omega} \, \times \, \vecbf{R} \right] \, p(\vecbf{\Omega}) \,
  p^{a}(\vecbf{R}) \, = \,
\sum_{m=-\infty}^{\infty} m \, p^{a}_{m,m} \, + \,
  2\pi \left[
    \int d^{2}\Omega \, \vecbf{\Omega} \, p(\vecbf{\Omega})
  \right] \, \times \,
  \left[
    \int d^{2}R \, \vecbf{R} \, p^{a}(\vecbf{R})
  \right] \, ,
\end{equation}
where $p^{a}(\vecbf{R})$ is the probability that a photon can pass through the
aperture within $d^{2}R$ of $\vecbf{R}$, and $p(\vecbf{\Omega})$ is the
probability that a photon arose from the celestial sphere within $d^{2}\Omega$
of $\vecbf{\Omega}$.  The source-dependent term is expressed as $\vecbf{\Omega}$
$\times$ $\vecbf{R}$ operating on the probabilities or the cross product of the
dipole moments of the probabilities.  I derive these equations and the variables
contained therein in Appendix \ref{App:IllForm}.

For an on-axis point source ($\rho$ $=$ $0$), the POAM and torque spectra in the
image plane are identical to the source-independent POAM spectra within the
aperture, or $\tilde{p}_{m,m}$ $=$ $\tau_{m,m}$ $=$ $p_{m,m}^{a}$.  Even if the
object under observation is not an on-axis point source, the ensemble of
$p_{m,m}^{a}$ and the POAM quantities formed from them represent reasonable
source-independent metrics.

The source-dependent terms, on the other hand, are dipole moments with $\pm$ $1$
selection rules, which means that they are identically zero on-axis and their
effects are relatively small off axis.  \citet{Elias08} called these terms
``pointing'' POAM or ``structure'' POAM.  The source structure cannot be easily
disentangled from the effects of propagation media and instrumentation.  They
are also zero when there is only a single non-zero $s_{k}(R)$.

\section{Interferometers} \label{Sec:Interferometer}

\citet{Elias08} derived POAM correlations and rancors for a single-baseline
optical interferometer.  They depend on baseline length, delay, telescope
aberrations, etc.  Since he was considering only a single observation with a
single pair of telescopes and integrating over the image plane, employing the
baseline vector instead of the two telescope vectors is acceptable.  For this
simple situation, the total POAM and torque can be set to zero because the
synthetic aperture origin can be always placed along the line containing the
single correlation measurement of two telescopes.

In this section, I derive image-plane POAM quantities for an interferometer.
There are slight differences between radio and optical interferometry.  In the
radio case, the electric fields between pairs of telescopes are multiplied and
averaged.  In the optical case, electric fields are summed, squared, and
averaged.  Mathematically, the same zero-spacing fluxes and visibilities can be
obtained in both the radio and optical.  I assume that the fringes are tracked
well enough to avoid any delay dependence.

\begin{figure}[!ht]
  \begin{center}
    \includegraphics[width=17cm,angle=0]{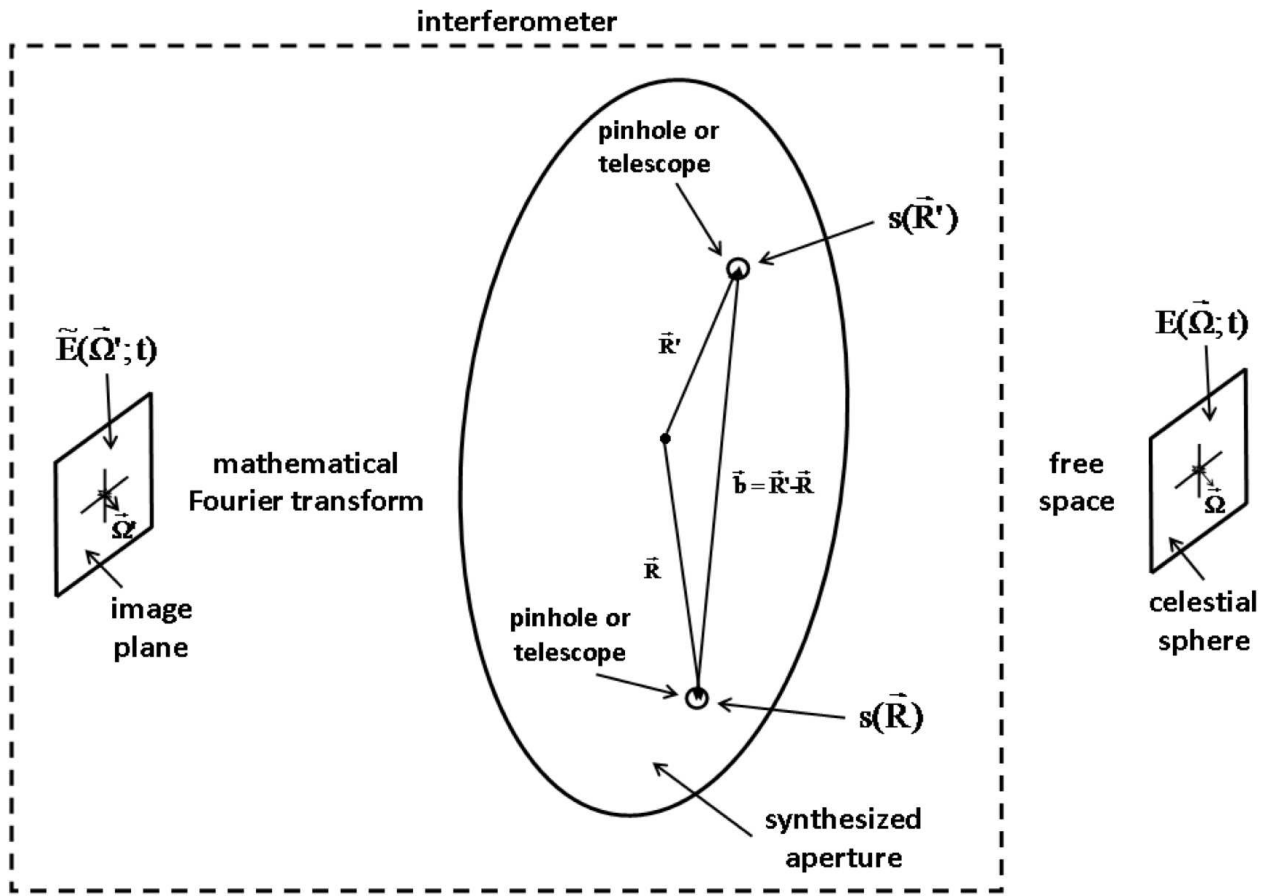}
  \end{center}
  \caption{Schematic diagram of an interferometer looking at a source on the
celestial sphere.  The coordinates are indicated and described in the text of
Sections \ref{Sec:Telescope} and \ref{Sec:Interferometer}.  The electric fields
(celestial sphere and image plane) and aperture function are also presented.
The aperture function refers to the synthesized aperture, not to the individual
telescope (or pinhole) apertures that sample it.}
  \label{Fig:Interferometer}
\end{figure}

Consider a single telescope behind an opaque aperture that contains a finite
number of imperfect pinholes with amplitude and phase errors.  The resulting
``dirty'' image is the convolution of the perfect diffraction-limited image and
the Fourier transform of the imperfect pinhole pattern, which is equivalent to
the ``optomechanical'' Fourier transform of the unnormalized visibilities from
all pinhole pairs.  An interferometer works in a similar manner.  Multiple
observations with an array of small telescopes, representing the imperfect
pinholes, form the ``synthesized'' aperture of a single large telescope (cf.
Figure \ref{Fig:Interferometer}).  Unnormalized visibilities of all telescope
pairs are then mathematically Fourier transformed to create the dirty image.
The sky-dependent response of the individual telescopes defines the image FOV
and their imperfections also affect the images (cf. Sections \ref{SSec:ApodTel}
and \ref{Sec:EVLA}).

\subsection{Initial Mathematics} \label{SSec:MathInt}

The single-telescope POAM mathematics of Section \ref{SSec:MathTel} can be used
directly with interferometers, taking into account the discrete sampling of the
aperture.  For the sake of illustration, however, I rewrite them in terms of
sampled unnormalized visibilities, which are the standard interferometer
observables.  The interferometer intensity response is
\begin{mathletters} \label{Ml:IntResponses}
\begin{eqnarray} \label{Eq:IntResponse}
\tilde{B}(\vecbf{\Omega}^{\prime}) &=
&\left<
  \frac{1}{2} \left| \tilde{E}(\vecbf{\Omega}^{\prime};t) \right|^{2}
\right> \, = \,
\left< \frac{1}{2}
  \left|
    \int d^{2}R^{\prime} \,
    e^{-j 2\pi \vecbf{R}^{\prime} \cdot \vecbf{\Omega}^{\prime}} \,
    s(\vecbf{R}^{\prime}) \, \mathcal{E}(\vecbf{R}^{\prime};t)
  \right|^{2}
\right> \nonumber \\ &=
&\int d^{2}R^{\prime} \, \int d^{2}R \,
  e^{-j 2\pi (\vecbf{R}^{\prime}-\vecbf{R}) \cdot \vecbf{\Omega}^{\prime}} \,
  \left[
    s(\vecbf{R}^{\prime}) \, s^{*}(\vecbf{R}) \,
    \mathcal{F}(\vecbf{R}^{\prime},\vecbf{R})
  \right] \, = \,
\int d^{2}R^{\prime} \, \int d^{2}R \,
  e^{-j 2\pi (\vecbf{R}^{\prime}-\vecbf{R}) \cdot \vecbf{\Omega}^{\prime}} \,
  \tilde{\mathcal{F}}(\vecbf{R}^{\prime},\vecbf{R}) \, , ~~~~~~~~
\end{eqnarray}
where $\mathcal{E}(\vecbf{R}^{\prime};t)$ is the electric field in the
synthesized aperture, $s(\vecbf{R}^{\prime})$ is the telescope-based gain
function of the synthesized aperture (analogous to the aperture function of
a single telescope, cf. Section \ref{Sec:Telescope}),
$\tilde{\mathcal{F}}(\vecbf{R}^{\prime},\vecbf{R})$ is the uncalibrated
unnormalized visibility, and
\begin{eqnarray} \label{Eq:SkyVis}
\mathcal{F}(\vecbf{R}^{\prime},\vecbf{R}) &=
&\mathcal{F}(\vecbf{R}^{\prime}-\vecbf{R}) \, = \,
\left<
  \frac{1}{2} \, \mathcal{E}(\vecbf{R}^{\prime};t) \,
  \mathcal{E}^{*}(\vecbf{R};t)
\right> \, = \,
\left<
  \frac{1}{2} \, \int d^{2}\Omega^{\prime\prime} \,
  e^{j 2\pi \vecbf{R}^{\prime} \cdot \vecbf{\Omega}^{\prime\prime}} \,
  E(\vecbf{\Omega}^{\prime\prime};t) \, \int d^{2}\Omega \,
  e^{-j 2\pi \vecbf{R} \cdot \vecbf{\Omega}} \, E^{*}(\vecbf{\Omega};t)
\right> \nonumber \\ &=
&\int d^{2}\Omega \,
  e^{j 2\pi (\vecbf{R}^{\prime}-\vecbf{R}) \cdot \vecbf{\Omega}} \,
  \left< \frac{1}{2} \left| E(\vecbf{\Omega};t) \right|^{2} \right> \, = \,
  \int d^{2}\Omega \,
  e^{j 2\pi (\vecbf{R}^{\prime}-\vecbf{R}) \cdot \vecbf{\Omega}} \,
  B(\vecbf{\Omega}) ~~~~~~
\end{eqnarray}
is the true unnormalized visibility under SAA.  Note that the image-plane
intensity of Equation \ref{Eq:IntResponse} is expressed in terms of two
aperture-plane integrals, as opposed to the standard single integral over
baseline $\vecbf{b}$ $=$ $\vecbf{R}^{\prime}$ $-$ $\vecbf{R}$ (the separation of
two telescopes), because POAM quantities depend on telescope position vectors
not baseline vectors.  In Appendix \ref{App:IntCorr}, I prove that this equation
can be converted to the baseline form used for non-POAM analysis.  I assume that
a single moving baseline produces all of the unnormalized visibilities.  All
formulae, however, can easily be generalized to multiple moving baselines.
\end{mathletters}

Expanding the exponential functions in Equation \ref{Eq:IntResponse} in terms
of Bessel functions, the $(m,m)^{th}$ POAM autocorrelation density becomes
\begin{mathletters}
\begin{eqnarray} \label{Eq:BTildeMM2}
\tilde{B}_{m,m}(\rho^{\prime}) \, = \,
\int d^{2}R^{\prime} \int d^{2}R \, J_{m}(2\pi R^{\prime} \rho^{\prime}) \,
  J_{m}(2\pi R \rho^{\prime}) \, e^{-jm (\psi^{\prime}-\psi)} \,
  \tilde{\mathcal{F}}(\vecbf{R}^{\prime},\vecbf{R}) \, . ~~~~~~
\end{eqnarray}
Integrating over the image plane, I obtain the $(m,m)^{th}$ POAM autocorrelation
\begin{equation} \label{Eq:BTildeMMTotal2}
\tilde{B}_{m,m} \, = \,
\lim_{\rho_{FOV} \rightarrow \infty} 2\pi \int_{0}^{\rho_{FOV}} d\rho^{\prime}
 \, \rho^{\prime} \, \tilde{B}_{m,m}(\rho^{\prime}) \, = \,
2\pi \int_{0}^{R_{int}} dR \, R ~ \frac{1}{2\pi} \int_{0}^{2\pi} d\psi^{\prime}
  ~ \frac{1}{2\pi} \int_{0}^{2\pi} d\psi ~ e^{-jm (\psi^{\prime} - \psi)} \,
  \tilde{\mathcal{F}}(\vecbf{r},\vecbf{R}) \, ,
\end{equation}
where $\vecbf{r}$ $=$ $(R \cos{\psi^{\prime}}, R \sin{\psi^{\prime}})$ and
$R_{int}$ is the radius of the synthesized aperture, both in units of
wavelength.  These equations are derived in Appendix \ref{App:IntCorr}.  As in
the single-telescope case (cf. Section \ref{SSec:MathTel}), Equation
\ref{Eq:BTildeMMTotal2} can be used to create all of the POAM quantities defined
in Section \ref{Sec:Definitions} as well as the illustrative forms of Section
\ref{SSec:IllForms} (cf. Appendix \ref{App:IntCorr}).  There are, however, two
complications.
\end{mathletters}

The only uncalibrated unnormalized visibilities that contribute to the $m^{th}$
POAM state autocorrelations are those which come from pairs of telescopes in the
same aperture ring $R$, i.e., pairs of telescopes that are the same distance
from the aperture reference point ($R$ $=$ $|\vecbf{r}|$ $=$
$|\vecbf{R}^{\prime}|$ $=$ $|\vecbf{R}|$, cf. Figure \ref{Fig:Interferometer}).
This result is not surprising given that POAM quantities are defined in rings.
As a matter of fact, it is possible to rewrite Equation \ref{Eq:BTildeMMTotal2}
as azimuthal Fourier series components of the azimuthal convolution of
uncalibrated aperture electric fields integrated over radius.  Unfortunately,
real interferometers do not have telescopes arranged in this manner, which means
that the true unnormalized visibilities and telescope-based gains must somehow
be interpolated onto a polar grid.

Single telescopes obtain data using an entire aperture.  Source-independent POAM
quantities calculated from an on-axis point source calibrator observation can be
used to judge the quality of separate science target observations if the
atmospheric statistics are $\approx$ consistent and the source structure does
not extend too far from the FOV center.  This strategy does not work for
interferometers.  They obtain data through a sampled synthetic aperture, so the
sample coverage for an on-axis point source calibrator and a science target will
be different.

Determining the optimum strategy to overcome these complications requires a
significant amount of effort.  Such work is beyond the scope of this paper, but
here I present two possible candidates that act as starting points for future
research.

When the ungridded discrete Fourier transform (DFT) of science target
uncalibrated unnormalized visibilities is calculated (no additional processing;
e.g., CLEAN), the resulting image is corrupted by incomplete sampling of the
synthetic aperture and gain errors.  If the inverse DFT (IDFT) of the corrupted
image is calculated on a polar grid, it effectively interpolates the
uncalibrated unnormalized visibilities so that they can be used directly in
Equation \ref{Eq:BTildeMM2} to form the POAM quantities of Sections
\ref{Sec:Definitions} and \ref{Sec:Telescope}.

Many interferometry imaging-processing techniques iteratively solve for sampled
true unnormalized visibilities and telescope-based gains while improving the
image model \citep{UrvashiIEEE,UrvashiPhD}. If the synthetic aperture is sampled
densely enough, the true unnormalized visibilities and telescope-based gains can
be interpolated onto a uniform polar grid so that the desired POAM quantities of
Sections \ref{Sec:Definitions} and \ref{Sec:Telescope} can be determined.  In
Appendix \ref{App:Interpolate}, I show that the interpolation kernel must be
azimuthally symmetric, or
$\mathcal{K}(\left|\vecbf{\mathcal{R}}-\vecbf{R}\right|)$, in order not to
further modulate the POAM spectrum.

These two strategies yield different results.  The DFT/IDFT interpolation method
includes the effects of both incomplete sampling and gain errors.  The
azimuthally symmetric interpolation kernel method, on the other hand, includes
only the effects of gain errors if the processing successfully removes artifacts
due to imperfect synthesized aperture sampling.  The DFT/IDFT interpolation
method includes pointing/structure POAM which cannot be easily disentangled from
source-independent terms (if the object under observation is an on-axis point
source, there is no pointing/structure POAM).  Conversely, the azimuthally
symmetric interpolation kernel method estimates the aperture functions, which
means that pointing/structure POAM can be disentangled from source-independent
terms.

\subsection{Telescope Apodization} \label{SSec:ApodTel}

The true unnormalized visibility (Equation \ref{Eq:SkyVis}) for a single
baseline, modified by the apodization of the individual telescopes, is
\begin{eqnarray} \label{Eq:SkyVisApod}
\mathcal{F}(\vecbf{R}^{\prime},\vecbf{R}) \, = \,
\int d^{2}\Omega \,
  e^{j 2\pi (\vecbf{R}^{\prime}-\vecbf{R}) \cdot \vecbf{\Omega}} \,
  \left[
    \mathcal{A}^{\prime}(\vecbf{\Omega}) \, \mathcal{A}^{*}(\vecbf{\Omega})
  \right] \, B(\vecbf{\Omega}) \, = \,
\int d^{2}\Omega \,
  e^{j 2\pi (\vecbf{R}^{\prime}-\vecbf{R}) \cdot \vecbf{\Omega}} \,
  \mathcal{P}(\vecbf{\Omega}) \, B(\vecbf{\Omega}) \, , ~~
\end{eqnarray}
where $\mathcal{A}^{\prime}(\vecbf{\Omega})$ and $\mathcal{A}(\vecbf{\Omega})$
are the complex electric-field sky-dependent gains of two telescopes at points
$\vecbf{R}^{\prime}$ and $\vecbf{R}$ in the observation plane, and
$\mathcal{P}(\vecbf{\Omega})$ is their combined power sky-dependent gain.  This
equation can easily be generalized for multiple moving baselines.  If the
electric-field gains are identical, the power gain is real.  If the converse is
true, the power gain is complex. A complex power gain represents a non-Hermitian
calibration error, $\mathcal{F}^{*}(\vecbf{R}^{\prime},\vecbf{R})$ $\neq$
$\mathcal{F}(\vecbf{R},\vecbf{R}^{\prime})$, which leads to a complex output
image (cf. Equation \ref{Eq:TelResponseA}).

Substituting Equation \ref{Eq:SkyVisApod} into Equations
\ref{Ml:IntResponses}a-b, I obtain a modified version of Equation
\ref{Eq:TelResponse}
\begin{equation} \label{Eq:TelResponseA}
\tilde{B}(\vecbf{\Omega}^{\prime}) \, = \,
  \int d^{2}\Omega \, P(\vecbf{\Omega}^{\prime},\vecbf{\Omega}) \,
  \mathcal{P}(\vecbf{\Omega}) \, B(\vecbf{\Omega}) \, .
\end{equation}
When expanded into POAM components, this Equation \ref{Eq:TelResponseA} becomes
\begin{mathletters} \label{Ml:ApodResponsePOAM}
\begin{equation} \label{Eq:TelResponseAPOAMSum}
\tilde{B}(\vecbf{\Omega}^{\prime}) \, = \,
\sum_{p=-\infty}^{\infty} \sum_{q=-\infty}^{\infty}
  \tilde{B}_{p,q}(\rho^{\prime}) \, e^{j(p-q)\phi^{\prime}} \, ,
\end{equation}
where the POAM correlations
\begin{equation} \label{Eq:TelResponseAPOAM}
\tilde{B}_{p,q}(\rho^{\prime}) \, = \,
\sum_{m=-\infty}^{\infty} \sum_{n=-\infty}^{\infty} 2\pi \int_{0}^{\infty}
d\rho \, \rho
\left[
  \sum_{k=-\infty}^{\infty} \sum_{l=-\infty}^{\infty}
  P_{p,q}^{-k,-l}(\rho^{\prime},\rho) \, \mathcal{P}_{k-l-m+n}(\rho)
\right] \, B_{m,n}(\rho)
\end{equation}
have an extra function
\begin{equation} \label{Eq:Pklmn}
\mathcal{P}_{k-l-m+n}(\rho) \, = \,
  \frac{1}{2\pi} \int_{0}^{2\pi} d\phi \, e^{-j(k-l-m+n)\phi} \,
  \mathcal{P}(\vecbf{\Omega}) \, .
\end{equation}
\end{mathletters}
The interferometric PSF input/output (separate) gain is
\begin{mathletters}
\begin{equation} \label{Eq:IOgainPSF}
P_{p,q}^{-k,-l}(\rho^{\prime},\rho) \, = \,
  D_{p}^{-k}(\rho^{\prime},\rho) \, D_{q}^{-l *}(\rho^{\prime},\rho) \, ,
\end{equation}
where
\begin{equation} \label{Eq:IOgainDF}
D_{m}^{-n}(\rho^{\prime},\rho) \, = \,
  \frac{1}{2\pi} \int_{0}^{2\pi} d\phi \, e^{jn \phi} \,
  D_{m}(\rho^{\prime},\vecbf{\Omega})
\end{equation}
is the interferometric DF input/output gain.  These gains are defined in Tables
2 and 4 of \citet{Elias08}.  I derive Equations \ref{Ml:ApodResponsePOAM}b-c in
Appendix \ref{App:ApodTel}.
\end{mathletters}

From Equation \ref{Eq:TelResponseAPOAM}, I see that sky-dependent gains do
indeed modulate POAM.  To understand these effects more clearly, I choose a
simple case where the interferometric synthetic aperture is fully sampled with
no amplitude or phase errors, which means that
\begin{mathletters}
\begin{equation} \label{Eq:Ppqpq}
P_{p,q}^{-k,-l}(\rho^{\prime},\rho) \, \rightarrow \,
  P_{p,q}^{-p,-q}(\rho^{\prime},\rho) \, \delta_{k,p} \, \delta_{l,q}
\end{equation}
\citep{Elias08}.  Equation \ref{Eq:TelResponseAPOAM} then becomes
\begin{equation} \label{Eq:TelResponseAPOAMpqpq}
\tilde{B}_{p,q}(\rho^{\prime}) \, = \,
\sum_{m=-\infty}^{\infty} \sum_{n=-\infty}^{\infty} 2\pi \int_{0}^{\infty}
d\rho \, \rho
\left[
  P_{p,q}^{-p,-q}(\rho^{\prime},\rho) \, \mathcal{P}_{p-q-m+n}(\rho)
\right] \, B_{m,n}(\rho) \, .
\end{equation}
Now only the sky-dependent gains distribute the input POAM correlation densities
to multiple output POAM correlation densities.  The index of the extra function
consists of the difference of image plane and celestial sphere index
differences.  To further verify that these mathematics are correct, I let the
sky-dependent gain exhibit only radial apodization, or
$\mathcal{P}(\vecbf{\Omega})$ $\rightarrow$ $\mathcal{P}(\rho)$.  The new extra
function
\end{mathletters}
\begin{mathletters}
\begin{equation} \label{Eq:Ppqmn}
\mathcal{P}_{p-q-m+n}(\rho) \, \rightarrow \,
  \mathcal{P}_{0}(\rho) \, \delta_{q,p-m+n}
\end{equation}
produces a one-to-one correspondence between input and output POAM correlation
densities
\begin{equation} \label{Eq:TelResponseAPOAMmn}
\tilde{B}_{m,n}(\rho^{\prime}) \, = \,
  2\pi \int_{0}^{\infty} d\rho \, \rho \,
  \left[
    \sum_{p=-\infty}^{\infty} P_{p,p-m+n}^{-p,-p+m-n}(\rho^{\prime},\rho)
  \right] \, \mathcal{P}_{0}(\rho) \, B_{m,n}(\rho) \, ,
\end{equation}
as expected.
\end{mathletters}

The sky-dependent gain for each antenna can be determined via holography, i.e.,
raster scans around a bright point source normalized (amplitude and phase) to a
reference antenna (cf. Section \ref{Sec:EVLA}).  The Fourier transforms of the
sky-dependent gains and their inverses are
\begin{eqnarray} \label{Eq:Holography}
a^{\prime}(\vecbf{r}^{\prime}) \, = \,
  \int d^{2}\Omega \, e^{-j 2\pi \vecbf{r}^{\prime} \cdot \vecbf{\Omega}} \,
  \mathcal{A}^{\prime}(\vecbf{\Omega})
  &~~~~ \stackrel{\mathcal{F}}{\Leftrightarrow} ~~~~&
\mathcal{A}^{\prime}(\vecbf{\Omega}) \, = \,
  \int d^{2}r^{\prime} \, e^{j 2\pi \vecbf{r}^{\prime} \cdot \vecbf{\Omega}} \,
  a^{\prime}(\vecbf{r}^{\prime}) \nonumber \\
a(\vecbf{r}) \, = \,
  \int d^{2}\Omega \, e^{-j 2\pi \vecbf{r} \cdot \vecbf{\Omega}} \,
  \mathcal{A}(\vecbf{\Omega})
  &~~~~ \stackrel{\mathcal{F}}{\Leftrightarrow} ~~~~&
\mathcal{A}(\vecbf{\Omega}) \, = \,
  \int d^{2}r \, e^{j 2\pi \vecbf{r} \cdot \vecbf{\Omega}} \, a(\vecbf{r}) \, ,
\end{eqnarray}
where $a^{\prime}(\vecbf{r}^{\prime})$ and $a(\vecbf{r})$ are the complex
holography functions representing aperture imperfections projected back to
planes in front of the telescopes.  The aperture coordinates are
$\vecbf{r}^{\prime}$ $=$
$(r^{\prime}\cos{\chi^{\prime}},r^{\prime}\sin{\chi^{\prime}})$ and $\vecbf{r}$
$=$ $(r\cos{\chi},r\sin{\chi})$.  These functions are conceptually identical to
the single-telescope aperture function (cf. Section \ref{SSec:MathTel}) and
interferometer synthesized aperture function (cf. Section \ref{SSec:MathInt}).
The POAM components of the inverse transforms are
\begin{mathletters}
\begin{eqnarray} \label{Eq:HolographyPOAM}
\mathcal{A}_{k}^{\prime}(\rho) \, = \,
  j^{k} \, 2\pi \int_{0}^{R_{tel}} dr^{\prime} \, r^{\prime} \,
  J_{k}(2\pi r^{\prime} \rho) \, a_{k}^{\prime}(r^{\prime})
  ~~~~~~ \mathrm{and} ~~~~~~
\mathcal{A}_{k}(\rho) \, = \,
  j^{k} \, 2\pi \int_{0}^{R_{tel}} dr \, r \, J_{k}(2\pi r \rho) \, a_{k}(r)
  \, ,
\end{eqnarray}
where
\begin{eqnarray} \label{Eq:AFourier}
\mathcal{A}_{k}^{\prime}(\rho) \, = \,
  \frac{1}{2\pi} \int_{0}^{2\pi} d\phi \, e^{-jk\phi} \,
  \mathcal{A}^{\prime}(\vecbf{\Omega})
  &~~~~ \stackrel{\mathcal{F}}{\Leftrightarrow} ~~~~&
  \mathcal{A}^{\prime}(\vecbf{\Omega}) \, = \,
  \sum_{k=-\infty}^{\infty} \mathcal{A}_{k}^{\prime}(\rho) \, e^{jk\phi}
  \nonumber \\
\mathcal{A}_{k}(\rho) \, = \,
  \frac{1}{2\pi} \int_{0}^{2\pi} d\phi \, e^{-jk\phi} \,
  \mathcal{A}(\vecbf{\Omega})
  &~~~~ \stackrel{\mathcal{F}}{\Leftrightarrow} ~~~~&
  \mathcal{A}(\vecbf{\Omega}) \, = \,
  \sum_{k=-\infty}^{\infty} \mathcal{A}_{k}(\rho) \, e^{jk\phi}
\end{eqnarray}
and
\begin{eqnarray} \label{Eq:aFourier}
a_{k}^{\prime}(r^{\prime}) \, = \,
  \frac{1}{2\pi} \int_{0}^{2\pi} d\chi^{\prime} \, e^{-jk\chi^{\prime}} \,
  a^{\prime}(\vecbf{r}^{\prime})
  &~~~~ \stackrel{\mathcal{F}}{\Leftrightarrow} ~~~~&
  a^{\prime}(\vecbf{r}^{\prime}) \, = \,
  \sum_{k=-\infty}^{\infty} a_{k}^{\prime}(r^{\prime}) \, e^{jk\chi^{\prime}}
  \nonumber \\
a_{k}(r) \, = \,
  \frac{1}{2\pi} \int_{0}^{2\pi} d\chi \, e^{-jk\chi} \,
  a(\vecbf{r})
  &~~~~ \stackrel{\mathcal{F}}{\Leftrightarrow} ~~~~&
  a(\vecbf{r}) \, = \,
  \sum_{k=-\infty}^{\infty} a_{k}(r) \, e^{jk\chi} \, .
\end{eqnarray}
Note that when the apertures are azimuthally symmetric, or
$a^{\prime}(\vecbf{r}^{\prime})$ $\rightarrow$ $a^{\prime}(r^{\prime})$ $=$
$a_{0}^{\prime}(r^{\prime})$ and $a(\vecbf{r})$ $\rightarrow$ $a(r)$ $=$
$a_{0}(r)$, only the $\mathcal{A}_{0}^{\prime}(\rho)$ and
$\mathcal{A}_{0}(\rho)$ terms are non-zero.  Further, if both antennas of a
baseline are azimuthally symmetric, it follows that their power pattern is also
azimuthally symmetric $\mathcal{P}(\vecbf{\Omega})$ $\rightarrow$
$\mathcal{P}(\rho)$ $=$ $\mathcal{P}_{0}(\rho)$ $=$
$\mathcal{A}_{0}^{\prime}(\rho)$ $\mathcal{A}_{0}^{*}(\rho)$ and does not
redistribute POAM states.
\end{mathletters}

\section{EVLA Holography and POAM} \label{Sec:EVLA}

Recent K-band ($\approx$ 24 GHz) holography observations of EVLA antennas were
processed during commissioning\footnote{Observing program THOL0001, source
3C273, 2011 October 14.} \citep{Brentjens}. The target is a bright point source.
Three of the antennas tracked the target and were used as amplitude and phase
references.  The rest of the antennas -- i.e., those under test -- executed
raster scans.  The resulting data were flagged, calibrated, and averaged before
calculating the Fourier transforms on Cartesian and polar output coordinates.
Calibration included removing the effects of pointing, focus, and subreflector
rotation offset, so the holography should $\sim$ represent the imperfections of
the antenna surfaces.

\begin{figure}[!ht]
  \begin{center}
    \includegraphics[width=15cm,angle=0]{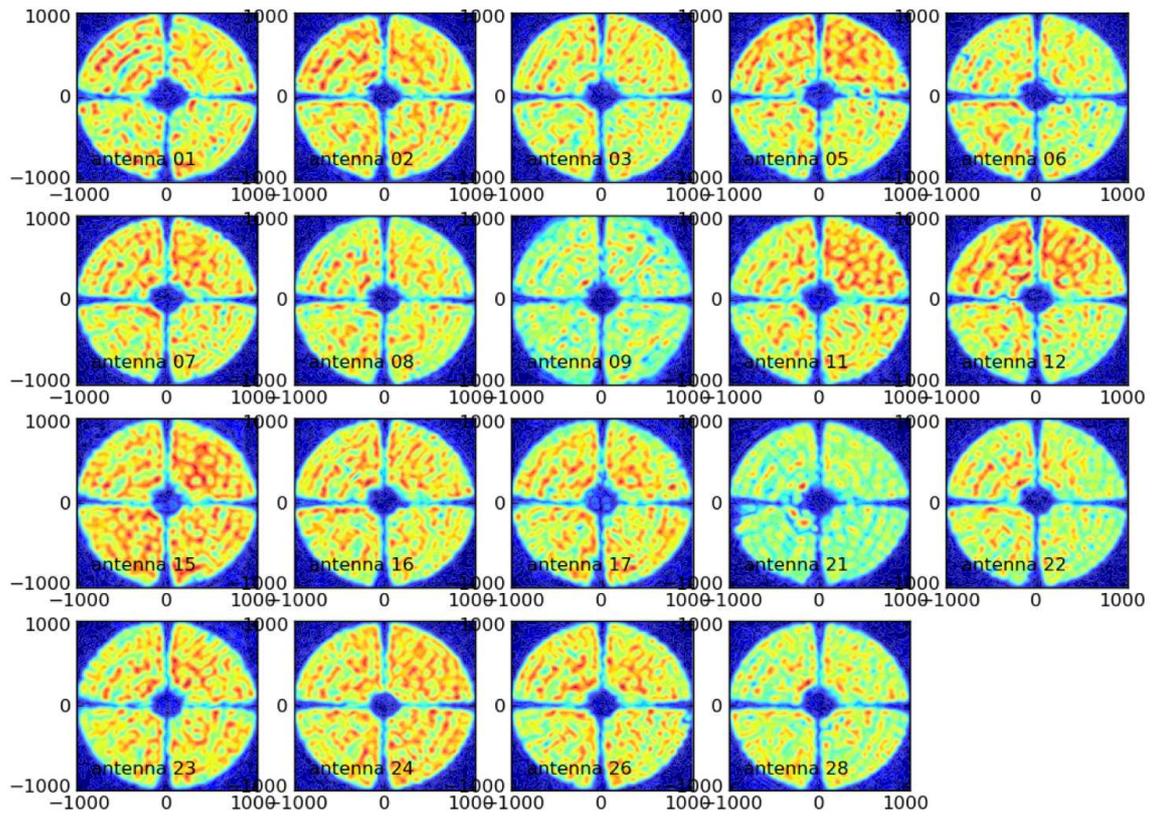}
  \end{center}
  \caption{Holographic amplitude measurements of the EVLA antennas.  The blue
and red colors represent the low and high reflectivity regions.  The coordinate
units are wavelengths.}
  \label{Fig:Amplitude_cg}
\end{figure}

\begin{figure}[!ht]
  \begin{center}
    \includegraphics[width=15cm,angle=0]{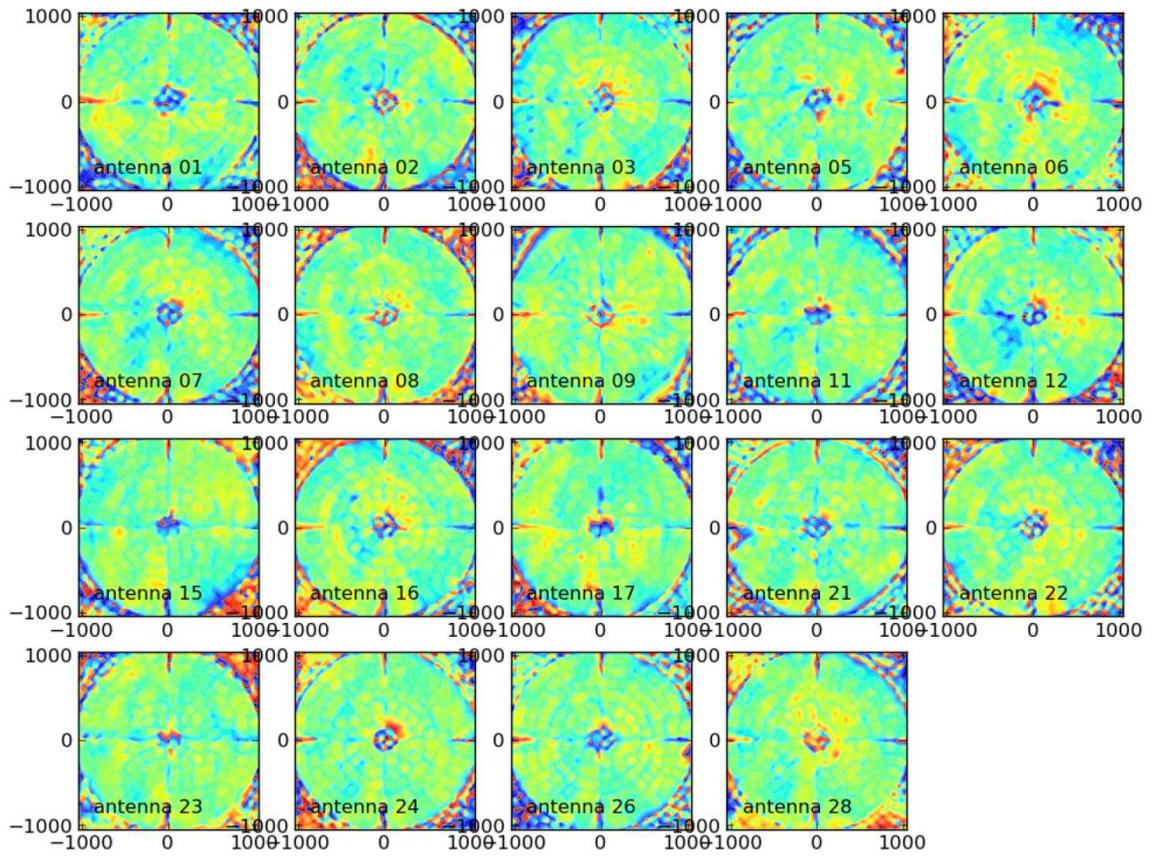}
  \end{center}
  \caption{Holographic phase measurements of the EVLA antennas.  The yellow
colors represent raised regions with respect to the fiducial dish surfaces.  The
RMS panel deviations are $\approx$ 200 $\mu$m.  The coordinate units are
wavelengths.}
  \label{Fig:Phase_cg}
\end{figure}

The amplitude and phase responses (Cartesian coordinates) are shown in Figures
\ref{Fig:Amplitude_cg} and \ref{Fig:Phase_cg}.  When calculating the Fourier
transforms, I oversampled the output by a factor of six for smoother
interpolation.  The circular shape of the dishes is clearly visible.  The
amplitude responses exhibit an opaque region near the center and four orthogonal
``spokes'' produced by the subreflector and its supports.  There are also clear
indications of both large and small spatial scale reflectivity features.  The
phase responses show random phases in the opaque regions, as expected.  The
locations of the small scale phase features (mottled patterns) are consistent
with individual misaligned panels on the reflector surface.  They tend to appear
in groups.

\begin{figure}[!ht]
  \begin{center}
    \includegraphics[width=15cm,angle=0]{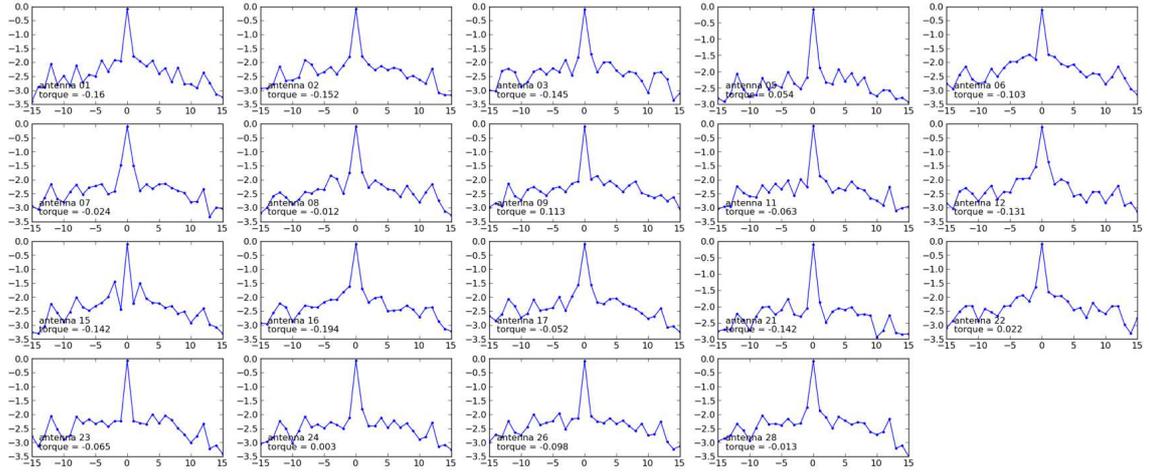}
  \end{center}
  \caption{Holographic probability spectra of the EVLA antennas.  The ordinates
are log (base 10) probability/100\% and the abscissae are POAM states. The total
torque is written on each subplot.}
  \label{Fig:POAMSpectrum}
\end{figure}

In Figure \ref{Fig:POAMSpectrum}, I display the sky-independent image-plane POAM
probability spectra (corresponding to the aperture POAM probability spectra in
Equations \ref{Eq:TelTorque1}, \ref{Eq:TelTorque2}, and \ref{Eq:TelTorque3}) as
well as the total torques.  There are no sky-dependent quantities (i.e., no
pointing/structure POAM) because the object under observation is an on-axis
point source.  To determine the sky-independent image-plane probabilities for
each telescope, I calculate the azimuthal Fourier series components of its
aperture function versus radius $a_{k}(\vecbf{r})$, form the squared magnitude
of each Fourier component versus radius $|a_{k}(\vecbf{r})|^{2}$, sum each
squared magnitude over radius to obtain $|a_{k}|^{2}$ (times $2 \pi$ $r$
$\Delta r$, where $\Delta r$ is the radial size of the aperture element), and
normalize the $|a_{k}|^{2}$ to the sum of the $|a_{k}|^{2}$ over $k$.  The range
of POAM components is limited to $\pm$ 15, which is $\approx$ the Nyquist limit
at the largest radius (I do not calculate higher POAM components even though
they are available because of oversampling).

When a perfect telescope observes an on-axis point source only the $n$ $=$ $0$
component of the image-plane POAM spectrum is non-zero.  Conversely, when an
imperfect telescope observes an on-axis point source the $n$ $=$ $0$ component
is reduced and the other components become non-zero.  Most of the $n$ $\neq$ $0$
probabilities are 1\% or less.  A few of the $n$ $=$ $\pm 1$ probabilities are
almost an order of magnitude larger, which could be caused by feed position
errors or uncalibrated pointing errors.  Summing over the $n$ $\neq$ $0$
probabilities, we find that $\sim$ 10\% of all photons are ``torqued'' away from
the $n$ $=$ $0$ state.  Also, note that most telescopes exhibit a non-zero total
torque but some (e.g., telescopes 24 and 28) exhibit $\approx$ zero total
torque (i.e., $\approx$ symmetric POAM spectra).

\section{Conclusion} \label{Sec:Conclusion}

With previously defined concepts and calculi \citep{Elias08}, I presented
generic expressions for POAM spectra, total POAM, torque spectra, and total
torque in the image plane.  I extended these functional forms to describe the
specific POAM behavior of both single telescopes and interferometers.  These
POAM quantities make excellent metrics, complimenting Zernike polynomials, for
describing the response of astronomical instruments.  Real holography
measurements of EVLA antennas demonstrated their utility.

Now that POAM metrics have been derived, it is incumbent on the author to make
them available to the astronomical community in an imaging package.  In the
future, I plan to extend them to handle spin-polarized and non-flat spectrum
sources.  Possible future studies include forming POAM quantities using real
interferometry visibility data, modeling instrumental aberrations and
turbulence of the troposphere/ionosphere in terms of POAM, POAM-based imaging
algorithms and constraints, POAM-based super resolution imaging
\citep{Tamburini} without interpolation or extrapolation, and POAM observations
of astrophysically important sources such as masers and black holes
\citep{Harwit03,TTM-TA}.

\begin{acknowledgements}
NME2 would like to thank Dr. Sanjay Bhatnagar for fruitful discussions and
advice, and Drs. Michiel Brentjens, Richard A. Perley, and Bryan J. Butler for
providing calibrated EVLA holography data.
\end{acknowledgements}

\newpage 

\newpage

\newpage

\newpage

\newpage

\newpage

\appendix

\section{POAM from the Celestial Sphere} \label{App:POAMCS}

\citet{Elias08} tacitly assumed that the total POAM on the celestial sphere was
zero for SAA sources and non-zero for non-SAA sources.  In this appendix, I
formally prove those statements for a single telescope, but the derivations
apply to all optical systems, including interferometers.

I consider the SAA case first.  If the instrument does not modulate the POAM
spectrum, the aperture function in Equation \ref{Eq:TelApFuncG} becomes
$s(\vecbf{R})$ $=$ $s_{0}(R)$.  The integral function in Equation
\ref{Eq:TelIntFunc} then simplifies to $\mathcal{J}_{p,q}(\rho^{\prime},\rho)$
$=$ $\mathcal{J}_{p,p}(\rho^{\prime},\rho)$ $\delta_{q,p}$.  When this result is
substituted into Equation \ref{Eq:TelPMM}, the $(m,m)^{th}$ POAM sensitivity is
no longer a function of $\phi$, or $P_{m,m}(\rho^{\prime},\vecbf{\Omega})$
$\rightarrow$ $P_{m,m}(\rho^{\prime},\rho)$ $=$
$\left|\mathcal{J}_{m,m}(\rho^{\prime},\rho)\right|^{2}$.  Upon inspection of
Equation \ref{Eq:TelIntFunc}, I find that
$\mathcal{J}_{-m,-m}(\rho^{\prime},\rho)$ $=$
$\mathcal{J}_{m,m}(\rho^{\prime},\rho)$, which means that
$P_{-m,-m}(\rho^{\prime},\rho)$ $=$ $P_{m,m}(\rho^{\prime},\rho)$.  Since
$P_{m,m}(\rho^{\prime},\rho)$ is even in $m$, the total image-plane POAM of
Equation \ref{Eq:TildeLz} is identically zero.  Because I initially assumed that
the instrument does not modulate POAM, it follows that the total POAM on the
celestial sphere must be zero as well.  Further, the total POAM on the celestial
sphere for SAA sources must always be zero because the behavior of the source
must be completely independent of the behavior of the instrument.
\textbf{Q.E.D.}

I now consider the non-SAA case using a completely spatially correlated source.
The spatial and temporal parts of the electric field on the celestial sphere
factor into separate functions, $E(\vecbf{\Omega};t)$ $=$ $E(\vecbf{\Omega})$
$f(t)$, where $f(t)$ is a random complex function.  The electric field in the
image plane (Equation \ref{Eq:TelResponse}) becomes
\begin{mathletters}
\begin{equation} \label{Eq:TelResponse2}
\tilde{E}(\vecbf{\Omega}^{\prime};t) \, = \,
\int d^{2}\Omega \, D(\vecbf{\Omega}^{\prime},\vecbf{\Omega}) \,
  E(\vecbf{\Omega};t) \, = \,
\left[
  \int d^{2}\Omega \, D(\vecbf{\Omega}^{\prime},\vecbf{\Omega}) \,
  E(\vecbf{\Omega})
\right] \, f(t) \, = \,
\tilde{E}(\vecbf{\Omega}^{\prime}) \, f(t) \, .
\end{equation}
Expanding the DF into sensitivities (Equation \ref{Eq:TelDM}) yields
\begin{equation} \label{Eq:TelResponse3}
\tilde{E}(\vecbf{\Omega}^{\prime};t) \, = \,
\sum_{m=-\infty}^{\infty} \tilde{E}_{m}(\rho^{\prime};t) \,
  e^{jm \phi^{\prime}} \, = \,
\sum_{m=-\infty}^{\infty} \left[ \tilde{E}_{m}(\rho^{\prime}) \, f(t) \right]
  \, e^{jm \phi^{\prime}} \, = \,
\sum_{m=-\infty}^{\infty} \left[
  \int d^{2}\Omega \, D_{m}(\rho^{\prime},\vecbf{\Omega}) \,
  E(\vecbf{\Omega}) \, f(t)
  \right] \, e^{jm \phi^{\prime}} \, . ~~~~~~
\end{equation}
Apart from the separation of the spatial and temporal components of the electric
field and the POAM states, these functional forms are identical to the SAA ones.
\end{mathletters}

With these POAM states, I form the image-plane POAM autocorrelation densities
(Equation \ref{Eq:BTildeMM}) and express the PSF sensitivities in terms of the
integral functions (Equation \ref{Eq:TelPMM})
\begin{mathletters}
\begin{equation} \label{Eq:BTildeMM6}
\tilde{B}_{m,m}(\rho^{\prime}) \, = \,
\left< \frac{1}{2} \left| \tilde{E}_{m}(\rho^{\prime};t) \right|^{2} \right>
  \, = \,
\left|
  \int d^{2}\Omega \, D_{m}(\rho^{\prime},\vecbf{\Omega}) \, E(\vecbf{\Omega})
\right|^{2} \, F \, = \,
\left|
  \sum_{k=-\infty}^{\infty} j^{k} \int d^{2}\Omega \,
  \mathcal{J}_{m,k}(\rho^{\prime},\rho) \, e^{-jk \phi} \, E(\vecbf{\Omega})
\right|^{2} \, F \, ,
\end{equation}
where $F$ $=$ $<\half |f(t)|^{2}>$ is the RMS of the random complex function.
Again I assume that the instrument does not modulate POAM, or
$\mathcal{J}_{m,k}(\rho^{\prime},\rho)$ $=$
$\mathcal{J}_{m,m}(\rho^{\prime},\rho)$ $\delta_{k,m}$, thereby collapsing the
sum in this equation.  Combining this fact with integration over radius
$\rho^{\prime}$ in the image plane, I obtain the total $(m,m)^{th}$ POAM
autocorrelation in the image plane
\begin{eqnarray} \label{Eq:BTildeMMTotal3}
\tilde{B}_{m,m} &=
&\lim_{\rho_{FOV} \rightarrow \infty} 2\pi \int_{0}^{\rho_{FOV}} d\rho^{\prime}
  \, \rho^{\prime} \tilde{B}_{m,m}(\rho^{\prime}) \nonumber \\ &=
&\lim_{\rho_{FOV} \rightarrow \infty} 2\pi \int_{0}^{\rho_{FOV}} d\rho^{\prime}
  \, \rho^{\prime} \left|
    \int d^{2}\Omega \,
    \mathcal{J}_{m,m}(\rho^{\prime},\rho) e^{-jm \phi} \, E(\vecbf{\Omega})
  \right|^{2} F =
\lim_{\rho_{FOV} \rightarrow \infty} 2\pi \int_{0}^{\rho_{FOV}} d\rho^{\prime}
  \, \rho^{\prime} \left|
    2\pi \int_{0}^{\infty} d\rho \, \rho
    \mathcal{J}_{m,m}(\rho^{\prime},\rho) E_{m}(\rho)
  \right|^{2} F \, , ~~~~~~~~~~~~
\end{eqnarray}
where
\begin{equation} \label{}
E_{m}(\rho) \, = \,
  \frac{1}{2\pi} \int_{0}^{2\pi} d\phi \, e^{-jm\phi} \, E(\vecbf{\Omega}) \, .
\end{equation}
This equation shows that $\tilde{B}_{-m,-m}$ $\neq$ $\tilde{B}_{m,m}$ because
$E_{-m}(\rho)$ $\neq$ $E_{m}(\rho)$ in general, which means that the total
image-plane POAM can be non-zero (Equation \ref{Eq:POAMTorque}).  The total
image-plane POAM is the same as the total source POAM because the instrument
does not modulate POAM, so if the former is non-zero the latter must be non-zero
as well.  \textbf{Q.E.D.}
\end{mathletters}

In the previous two proofs, I solved for the POAM autocorrelations in the image
plane and inferred the form for the POAM autocorrelations on the celestial
sphere.  Here I provide a bonus proof employing some mathematics that are not
found elsewhere in this paper.  I only use the POAM autocorrelations on the
celestial sphere and I express them in terms of azimuthal correlations of
electric fields.  This proof can be used to describe the POAM behavior of SAA
and non-SAA sources.

The $(m,m)^{th}$ POAM autocorrelation on the celestial sphere is
\begin{mathletters}
\begin{equation} \label{Eq:A.3a}
B_{m,m} \, = \,
2\pi \int_{0}^{\infty} d\rho \, \rho \, B_{m,m}(\rho) \, = \,
2\pi \int_{0}^{\infty} d\rho \, \rho \,
  \left< \frac{1}{2} \left| E_{m}(\rho;t) \right|^{2} \right> \, = \,
\frac{1}{2\pi} \int_{0}^{2\pi} d\phi^{\prime} \, e^{-j m \phi^{\prime}}
  \frac{1}{2\pi} \int_{0}^{2\pi} d\phi \, e^{j m \phi} \,
  B(\phi^{\prime},\phi) \, ,
\end{equation}
where
\begin{equation} \label{Eq:A.3b}
B(\phi^{\prime},\phi) \, = \,
  2\pi \int_{0}^{\infty} d\rho \, \rho \, B(\rho,\phi^{\prime},\phi)
\end{equation}
is the temporal correlation of the electric fields on the celestial sphere
between azimuths $\phi^{\prime}$ and $\phi$ integrated over radius $\rho$, and
\begin{eqnarray} \label{Eq:A.3c}
B(\rho,\phi^{\prime},\phi) &=
&\left<
  \frac{1}{2} E(\rho,\phi^{\prime};t) \, E^{*}(\rho,\phi;t)
\right> \, = \,
  \left<\frac{1}{2}
    \left[ E_{r}(\rho,\phi^{\prime};t) + j E_{i}(\rho,\phi^{\prime};t) \right]
    \left[ E_{r}(\rho,\phi;t) - j E_{i}(\rho,\phi;t) \right]
  \right> \nonumber \\ &=
&\left[
  \left<\frac{1}{2} E_{r}(\rho,\phi^{\prime};t) \, E_{r}(\rho,\phi;t)\right> +
  \left<\frac{1}{2} E_{i}(\rho,\phi^{\prime};t) \, E_{i}(\rho,\phi;t)\right>
\right] \, + \,
j \left[
  -\left<\frac{1}{2} E_{r}(\rho,\phi^{\prime};t) \, E_{i}(\rho,\phi;t)\right> +
  \left<\frac{1}{2} E_{i}(\rho,\phi^{\prime};t) \, E_{r}(\rho,\phi;t)\right>
\right] \nonumber \\ &=
&\left[
  B_{rr}(\rho,\phi^{\prime},\phi) + B_{ii}(\rho,\phi^{\prime},\phi)
\right] \, + \,
j \left[
  - B_{ri}(\rho,\phi^{\prime},\phi) + B_{ir}(\rho,\phi^{\prime},\phi)
\right] \, = \,
  B_{r}(\rho,\phi^{\prime},\phi) \, + \, j B_{i}(\rho,\phi^{\prime},\phi)
\end{eqnarray}
is the temporal correlation of the electric fields on the celestial sphere
between azimuths $\phi^{\prime}$ and $\phi$ at radius $\rho$.  In general,
$B(\phi^{\prime},\phi)$ and $B(\rho,\phi^{\prime},\phi)$ are complex numbers
written in terms of complex electric fields.  Similarly, the $(-m,-m)^{th}$ POAM
autocorrelation on the celestial sphere is
\begin{equation} \label{Eq:A.3d}
B_{-m,-m} \, = \,
2\pi \int_{0}^{\infty} d\rho \, \rho \, B_{-m,-m}(\rho) \, = \,
2\pi \int_{0}^{\infty} d\rho \, \rho \,
  \left< \frac{1}{2} \left| E_{-m}(\rho;t) \right|^{2} \right> \, = \,
\frac{1}{2\pi} \int_{0}^{2\pi} d\phi^{\prime} \, e^{-j m \phi^{\prime}}
  \frac{1}{2\pi} \int_{0}^{2\pi} d\phi \, e^{j m \phi} \,
  B^{*}(\phi^{\prime},\phi) \, ,
\end{equation}
It is obtained by replacing $m$ $\rightarrow$ $-m$ and exchanging
$\phi^{\prime}$ $\leftrightarrow$ $\phi$ in Equation \ref{Eq:A.3a}.
\end{mathletters}

As stated elsewhere, $l_{Z}$ $=$ $0$ when $B_{-m,-m}$ $=$ $B_{m,m}$ for all $m$.
I define the quantity
\begin{mathletters}
\begin{equation} \label{Eq:A.4a}
\Delta B_{m} \, = \,
B_{m,m} \, - \, B_{-m,-m} \, = \,
j2 \, \frac{1}{2\pi} \int_{0}^{2\pi} d\phi^{\prime} \, e^{-j m \phi^{\prime}}
  \frac{1}{2\pi} \int_{0}^{2\pi} d\phi \, e^{j m \phi} \,
  B_{i}(\phi^{\prime},\phi) \, ,
\end{equation}
where
\begin{equation} \label{Eq:A.4b}
B_{i}(\phi^{\prime},\phi) \, = \,
2\pi \int_{0}^{\infty} d\rho \, \rho \, B_{i}(\rho,\phi^{\prime},\phi) \, = \,
-2\pi \int_{0}^{\infty} d\rho \, \rho \, B_{ri}(\rho,\phi^{\prime},\phi) \, + \,
  2\pi \int_{0}^{\infty} d\rho \, \rho \, B_{ir}(\rho,\phi^{\prime},\phi)
  \, = \,
-B_{ri}(\phi^{\prime},\phi) \, + \, B_{ir}(\phi^{\prime},\phi) \, .
\end{equation}
For an SAA source Equations \ref{Eq:A.3c} and \ref{Eq:A.4b} lead to
$B_{ri}(\phi^{\prime},\phi)$ $=$ $B_{ir}(\phi^{\prime},\phi)$,
$B_{i}(\phi^{\prime},\phi)$ $=$ $0$, $\Delta B_{m}$ $=$ $0$ for all $m$, and
thus $l_{Z}$ $=$ $0$. In other words, if the complex electric fields are
spatially uncorrelated, their real and imaginary parts are spatially
uncorrelated as well.  Conversely, for a non-SAA source I find that
$B_{ri}(\phi^{\prime},\phi)$ $\neq$ $B_{ir}(\phi^{\prime},\phi)$,
$B_{i}(\phi^{\prime},\phi)$ $\neq$ $0$, $\Delta B_{m}$ $\neq$ $0$ for all $m$,
and thus $l_{Z}$ $\neq$ $0$.  \textbf{Q.E.D.}
\end{mathletters}

\section{Telescope POAM Autocorrelations} \label{App:TelCorr}

The $(m,m)^{th}$ single-telescope PSF sensitivity (Equation \ref{Eq:TelPMM}) is
the squared magnitude of the $m^{th}$ single-telescope DF sensitivity, or
$P_{m,m}(\rho^{\prime},\vecbf{\Omega})$ $=$
$\left|D_{m}(\rho^{\prime},\vecbf{\Omega})\right|^{2}$.  The $m^{th}$
single-telescope DF sensitivity is simply the azimuthal Fourier component in the
image plane of the DF, or
\begin{equation} \label{Eq:TelDM2}
D_{m}(\rho^{\prime},\vecbf{\Omega}) \, = \,
\frac{1}{2\pi} \, \int_{0}^{2\pi} \, d\phi^{\prime} \, e^{-j m \phi^{\prime}}
  \, D(\vecbf{\Omega}^{\prime},\vecbf{\Omega}) \, = \,
\int d^{2}R \, 
  \left[
    \frac{1}{2\pi} \int_{0}^{2\pi} d\phi^{\prime} \, e^{-jm \phi^{\prime}} \,
    e^{-j2\pi \vecbf{R} \cdot \vecbf{\Omega}^{\prime}} \,
  \right] \, e^{j2\pi \vecbf{R} \cdot \vecbf{\Omega}} \, s(\vecbf{R}) .
\end{equation}
The quantity in square brackets is $j^{-m}$ $J_{m}(2\pi R \rho^{\prime})$
$e^{-jm \psi}$.  When the aperture function is expanded into azimuthal Fourier
components, I obtain
\begin{eqnarray} \label{Eq:TelDM3}
D_{m}(\rho^{\prime},\vecbf{\Omega}) =
  j^{-m} \sum_{p=-\infty}^{\infty} 2\pi \int_{0}^{R_{tel}} dR \, R \,
  J_{m}(2\pi R \rho^{\prime}) \left[
    \frac{1}{2\pi} \int_{0}^{2\pi} d\psi \, e^{j(p-m) \psi}
    e^{j2\pi \vecbf{R} \cdot \vecbf{\Omega}} 
  \right] s_{p}(R) \, . ~~~~
\end{eqnarray}
The quantity in square brackets is $j^{p-m}$ $J_{p-m}(2\pi R \rho)$ $e^{j(p-m)
\phi}$.  After replacing $p$ with $m-k$ and rearranging, the result is
\begin{equation} \label{Eq:TelDM4}
D_{m}(\rho^{\prime},\vecbf{\Omega}) \, = \,
j^{-m} \, \sum_{k=-\infty}^{\infty} \, j^{k} \, \left[
  2\pi \, \int_{0}^{R_{tel}} \, dR \, R \, J_{m}(2\pi R \rho^{\prime}) \,
  s_{m-k}(R) \, J_{k}(2\pi R \rho)
  \right] \, e^{-jk \phi} \, = \,
j^{-m} \, \sum_{k=-\infty}^{\infty} \, j^{k} \,
  \mathcal{J}_{m,k}(\rho^{\prime},\rho) \, e^{-jk \phi} \, .
\end{equation}
The squared magnitude of this equation is identical to Equation
\ref{Eq:TelPMMTotal}, which can be substituted into Equation \ref{Eq:BTildeMM}
to give the complete $(m,m)^{th}$ single-telescope POAM autocorrelation
densities.
\textbf{Q.E.D.}

\section{Illustrative Forms} \label{App:IllForm}

To derive the first illustrative form, I change the indices $k$ $\rightarrow$
$m$ $-$ $p$ and $l$ $\rightarrow$ $m$ $-$ $q$ of Equations \ref{Eq:TelPMMTotal}
and \ref{Eq:TildeLz}.  The resulting total POAM kernel becomes
\begin{mathletters} \label{Ml:TorqueKernel}
\begin{equation} \label{Eq:TorqueKernel}
\tilde{\mathcal{L}}_{Z}(\vecbf{\Omega}) \, = \,
  \sum_{p=-\infty}^{\infty} \sum_{q=-\infty}^{\infty} j^{-(p-q)} \,
  2\pi \int_{0}^{R_{tel}} dR \, R \, s_{p}(R) \, s^{*}_{q}(R) \,
  \mathcal{M}_{p,q}(2\pi R \rho) \, e^{j(p-q) \phi} \, ,
\end{equation}
where
\begin{equation} \label{Eq:MPQ}
\mathcal{M}_{p,q}(2\pi R \rho) \, = \,
\sum_{m=-\infty}^{\infty} m \, J_{m-p}(2\pi R \rho) \,
  J_{m-q}(2\pi R \rho) \, = \,
p \, \delta_{q,p} \, + \, \frac{1}{2} \left[ 2\pi R \rho \right] \,
  \delta_{q,p+1} \, + \, \frac{1}{2} \left[ 2\pi R \rho \right] \,
  \delta_{q,p-1} \, .
\end{equation}
When Equations \ref{Ml:TorqueKernel}a-b are substituted back into Equation
\ref{Eq:TildeLz}, I obtain
\end{mathletters}
\begin{mathletters} \label{Ml:Ill1}
\begin{equation} \label{Eq:Torque2}
\tilde{l}_{Z} \, = \,
  \sum_{m=-\infty}^{\infty} m \frac{S_{m,m} \, B}{\tilde{B}} +
  j \frac{1}{2} \, 2\pi \int_{0}^{\infty} d\rho \, \rho \,
  \mathcal{D}_{+}(\rho) \mathcal{B}_{+1}(\rho) / \tilde{B} -
  j \frac{1}{2} \, 2\pi \int_{0}^{\infty} d\rho \, \rho \, \mathcal{D}_{-}(\rho)
  \mathcal{B}_{-1}(\rho) / \tilde{B} \, , ~~~~
\end{equation}
where
\begin{equation} \label{Eq:SMM}
S_{m,m} \, = \,
  2\pi \int_{0}^{R_{tel}} dR \, R \, S_{m,m}(R) \, = \,
  2\pi \int_{0}^{R_{tel}} dR \, R \, \left| s_{m}(R) \right|^{2}
\end{equation}
is the integrated and squared $(m,m)^{th}$ component of the aperture function,
the
\begin{equation} \label{Eq:Dpm}
\mathcal{D}_{\mp}(\rho) \, = \,
  2\pi \int_{0}^{R_{tel}} dR \, R \, \left( 2\pi R \rho \right) \,
  \left[ \sum_{m=-\infty}^{\infty} s_{m}(R) \, s^{*}_{m \mp 1}(R) \right]
\end{equation}
are the aperture dipole-moment functions, and the
\begin{equation} \label{Eq:Rancor_pm_1}
\mathcal{B}_{\mp 1}(\rho) \, = \,
\frac{1}{2\pi} \int_{0}^{2\pi} d\phi \, e^{\pm j \phi} \, B(\vecbf{\Omega})
  \, = \,
\frac{1}{2\pi} \int_{0}^{2\pi} d\phi \, e^{\pm j \phi}
  \left[
    \sum_{n=-\infty}^{\infty} \sum_{m=-\infty}^{\infty}
    B_{n,m}(\rho) \, e^{j (n-m) \phi}
  \right] \, = \,
  \sum_{n=-\infty}^{\infty} B_{n,n \pm 1}(\rho)
\end{equation}
are the first-order source rancors \citep[Equations 12a-b]{Elias08}.  Note that
$p$ was replaced with $m$ so that Equations \ref{Ml:Ill1}a-c are consistent with
the equations in Section \ref{SSec:IllForms}.  The total intensity in the image
plane can be rewritten as
\end{mathletters}
\begin{mathletters} \label{Ml:AllBTilde2-3}
\begin{equation} \label{Eq:AllBTilde2}
\tilde{B} \, = \,
\lim_{\rho_{FOV} \rightarrow \infty} \int_{\rho_{FOV}} d^{2}\Omega^{\prime}
  \tilde{B}(\vecbf{\Omega}^{\prime}) \, = \,
\int d^{2}\Omega \int d^{2}R \int d^{2}R^{\prime}
  \left[
    \lim_{\rho_{FOV} \rightarrow \infty} \int_{\rho_{FOV}} d^{2}\Omega^{\prime}
    e^{-j2\pi (\vecbf{R} - \vecbf{R}^{\prime}) \cdot \vecbf{\Omega}^{\prime}}
  \right]
  e^{j2\pi (\vecbf{R} - \vecbf{R}^{\prime}) \cdot \vecbf{\Omega}}
  s(\vecbf{R}) s^{*}(\vecbf{R}^{\prime}) B(\vecbf{\Omega}) \, .
\end{equation}
The quantity in the square brackets approaches
$\delta(\vecbf{R} - \vecbf{R}^{\prime})$, which means that
\begin{equation} \label{Eq:AllBTilde3}
\tilde{B} \, \approx \,
  \left[ \int d^{2}R \, S(\vecbf{R}) \right] \,
  \left[ \int d^{2}\Omega \, B(\vecbf{\Omega}) \right] \, = \,
  S \, B \, ,
\end{equation}
where $S(\vecbf{R})$ $=$ $|s(\vecbf{R})|^{2}$.  When this equation is
substituted into Equation \ref{Eq:Torque2}, I can define the aperture
probabilities $p^{a}_{m,m}$ $=$ $S_{m,m}$ $/$ $S$, the aperture transitional
probability densities $p^{a}_{m,m \mp 1}(R)$ $=$ $s_{m}(R)$ $s^{*}_{m \mp 1}(R)$
$/$ $S$, and the celestial sphere transitional probability densities
$p_{n,n \pm 1}(\rho)$ $=$ $B_{n,n \pm 1}(\rho)$ $/$ $B$.  Keeping in mind that
$\mathcal{B}_{-1}(\rho)$ $=$ $\mathcal{B}_{+1}^{*}(\rho)$ and
$\mathcal{D}_{-}(\rho)$ $=$ $\mathcal{D}_{+}^{*}(\rho)$, the result can be
rearranged to obtain Equation \ref{Eq:TelTorque1}.  \textbf{Q.E.D.}
\end{mathletters}

To derive the second illustrative form, I rewrite the sum over transitional
probabilities in the aperture in Equation \ref{Eq:TelTorque1} as
\begin{mathletters} \label{Ml:ssSquare}
\begin{equation} \label{Eq:ssSquare}
\sum_{m=-\infty}^{\infty} p^{a}_{m,m \mp 1}(R) \, = \,
\sum_{m=-\infty}^{\infty} \frac{s_{m}(R) \, s^{*}_{m \mp 1}(R)}{S} \, = \,
\frac{1}{2\pi} \int_{0}^{2\pi} d\psi \, \frac{1}{2\pi} \int_{0}^{2\pi}
  d\psi^{\prime} \, e^{\mp j \psi^{\prime}}
  \left(
    \sum_{m=-\infty}^{\infty} e^{-jm (\psi - \psi^{\prime})}
  \right) \, \frac{s(\vecbf{R}) \, s^{*}(\vecbf{r})}{S} \, ,
\end{equation}
where $\vecbf{R}$ $=$ $(R\cos{\psi},R\sin{\psi})$ and $\vecbf{r}$ $=$
$(R\cos{\psi^{\prime}},R\sin{\psi^{\prime}})$ (both aperture points are located
in the same ring $R$ $=$ $|\vecbf{R}|$ $=$ $|\vecbf{r}|$).  The quantity in
parentheses is $2\pi$ $\delta(\psi^{\prime} - \psi)$, so
\begin{equation} \label{Eq:ssSquare2}
\sum_{m=-\infty}^{\infty} p^{a}_{m,m \mp 1}(R) \, = \,
\sum_{m=-\infty}^{\infty} \frac{s_{m}(R) \, s^{*}_{m \mp 1}(R)}{S} \, = \,
\frac{1}{2\pi} \int_{0}^{2\pi} d\psi \, e^{\mp j \psi} \,
  \frac{\left| s(\vecbf{R}) \right|^{2}}{S} \, = \,
\frac{1}{2\pi} \int_{0}^{2\pi} d\psi \, e^{\mp j \psi} \,
  \frac{S(\vecbf{R})}{S} \, = \,
\frac{\mathcal{S}_{\pm 1}(R)}{S} \, ,
\end{equation}
which are the normalized first-order aperture rancor gains.  Substituting
Equation \ref{Eq:ssSquare2} into Equation \ref{Eq:Dpm}, Equation
\ref{Eq:Torque2} can be rearranged to form Equation \ref{Eq:TelTorque2}.
\textbf{Q.E.D.}
\end{mathletters}

To derive the third illustrative form, the imaginary part of the integrand of
Equation \ref{Eq:TelTorque2} can be simplified to
\begin{eqnarray} \label{Eq:ImPart}
\mathrm{Im} \left( 2\pi R \rho \right)
  \left[
    \frac{\mathcal{B}_{\mp 1}(\rho)}{B}
  \right]
  \left[
    \frac{\mathcal{S}_{\pm 1}(R)}{S}
  \right] &=
&\pm \, \frac{1}{2\pi} \int_{0}^{2\pi} d\phi \, \frac{1}{2\pi} \int_{0}^{2\pi}
  d\psi ~ 2\pi \left[ R \rho \sin{(\phi-\psi)} \right]
  \frac{B(\vecbf{\Omega})}{B} \, \frac{S(\vecbf{R})}{S} \nonumber \\ &=
&\pm \frac{1}{2\pi} \int_{0}^{2\pi} d\phi \, \frac{1}{2\pi} \int_{0}^{2\pi}
  d\psi ~ 2\pi \left[ \vecbf{\Omega} \times \vecbf{R} \right]
  \frac{B(\vecbf{\Omega})}{B} \, \frac{S(\vecbf{R})}{S} \, .
\end{eqnarray}
When this expression is substituted back into Equation \ref{Eq:TelTorque2},
Equation \ref{Eq:TelTorque3} is obtained.  \textbf{Q.E.D.}

\section{Interferometer POAM Autocorrelations} \label{App:IntCorr}

To prove that Equation \ref{Eq:IntResponse} can be expressed in terms of a
single integral over baseline for standard non-POAM interferometry analysis, I
substitute $\vecbf{b}$ $=$ $\vecbf{R}^{\prime}$ $-$ $\vecbf{R}$ to obtain
\begin{equation} \label{Eq:IntResponse2}
\tilde{B}(\vecbf{\Omega}^{\prime}) =
\int d^{2}R \, \int_{+\vecbf{R}} d^{2}b
  e^{-j 2\pi \vecbf{b} \cdot \vecbf{\Omega}^{\prime}}
  s(\vecbf{R}+\vecbf{b}) s^{*}(\vecbf{R}) \mathcal{F}(\vecbf{b}) =
\int d^{2}b \, e^{-j 2\pi \vecbf{b} \cdot \vecbf{\Omega}^{\prime}}
  \left[
    \int d^{2}R \, s(\vecbf{R}+\vecbf{b}) s^{*}(\vecbf{R})
  \right] \mathcal{F}(\vecbf{b}) =
  \int d^{2}b \, e^{-j 2\pi \vecbf{b} \cdot \vecbf{\Omega}^{\prime}}
  S(\vecbf{b}) \mathcal{F}(\vecbf{b}) , ~
\end{equation}
where $S(\vecbf{b})$ is the baseline-based gain function.  The $+\vecbf{R}$
shift in the $\vecbf{b}$ integral can be ignored.  The integral inside the
square brackets must be the baseline-based gain function.  Each point in the
aperture function $s(\vecbf{R})$ corresponds to the position of a telescope.
The aperture function appears twice in the integral because there are two
telescopes in a baseline.  The two telescopes are separated by the baseline.
\textbf{Q.E.D.} 

To derive the $(m,m)^{th}$ interferometer POAM autocorrelation density, I
substitute
\begin{mathletters}
\begin{equation} \label{Eq:JExp}
e^{j 2\pi \vecbf{R} \cdot \vecbf{\Omega}} \, = \,
e^{j 2\pi R \rho \cos{(\psi - \phi)}} \, = \,
  \sum_{k=-\infty}^{\infty} j^{k} \, J_{k}(2\pi R \rho) \,
  e^{j k (\psi-\phi)}
\end{equation}
into Equation \ref{Eq:IntResponse} and find that
\begin{eqnarray} \label{Eq:IntResponse3}
\tilde{B}(\vecbf{\Omega}^{\prime}) &=
&\int d^{2}R^{\prime} \int d^{2}R
  \left[
    \sum_{m=-\infty}^{\infty} j^{-m} J_{m}(2\pi R^{\prime} \rho^{\prime})
    e^{-j m (\psi^{\prime} - \phi^{\prime})}
  \right]
  \left[
    \sum_{n=-\infty}^{\infty} j^{n} J_{n}(2\pi R \rho^{\prime})
    e^{j n (\psi - \phi^{\prime})}
  \right] \tilde{\mathcal{F}}(\vecbf{R}^{\prime},\vecbf{R}) \nonumber \\ &=
&\sum_{m=-\infty}^{\infty} \sum_{n=-\infty}^{\infty}
  \left[
    j^{-(m-n)} \int d^{2}R^{\prime} \int d^{2}R
    J_{m}(2\pi R^{\prime} \rho^{\prime}) J_{n}(2\pi R \rho^{\prime})
    e^{-j m \psi^{\prime}} e^{j n \psi}
    \tilde{\mathcal{F}}(\vecbf{R}^{\prime},\vecbf{R})
  \right] e^{j (m-n) \phi^{\prime}} \nonumber \\ &=
&\sum_{m=-\infty}^{\infty} \sum_{n=-\infty}^{\infty}
  \tilde{B}_{m,n}(\rho^{\prime}) \, e^{j (m-n) \phi^{\prime}} \, .
\end{eqnarray}
Each $m$ $=$ $n$ term is identical to Equation \ref{Eq:BTildeMM2}.  When
Equation \ref{Eq:BTildeMM2} is integrated over radius in the image plane,
this integral
\begin{equation} \label{Eq:Collapse}
\lim_{\rho_{FOV} \rightarrow \infty} 2\pi \int_{0}^{\rho_{FOV}} d\rho^{\prime}
  \, \rho^{\prime} \, J_{m}(2\pi R^{\prime} \rho^{\prime}) \,
  J_{m}(2\pi R \rho^{\prime}) \, = \,
  \frac{\delta( R^{\prime} - R )}{2\pi R^{\prime}}
\end{equation}
appears as one of the factors in the result.  This Dirac delta function
collapses the $R^{\prime}$ integral, leaving Equation \ref{Eq:BTildeMMTotal2}.
\textbf{Q.E.D.}
\end{mathletters}

To prove that the interferometer POAM autocorrelations of Equation
\ref{Eq:BTildeMMTotal2} can be converted the single-telescope illustrative forms
(cf. Section \ref{SSec:IllForms}), I form the total POAM in the image plane
\begin{mathletters}
\begin{equation} \label{Eq:BTildeMMTotal2Summed}
\tilde{l}_{Z} \, = \,
\sum_{m=-\infty}^{\infty} m \, \tilde{p}_{m,m} \, = \,
  \frac{1}{\tilde{B}} \sum_{m=-\infty}^{\infty} m \, \tilde{B}_{m,m} \, = \,
\frac{1}{\tilde{B}} \,
  2\pi \int_{0}^{R_{int}} dR \, R \frac{1}{2\pi} \int_{0}^{2\pi} d\psi^{\prime}
  \frac{1}{2\pi} \int_{0}^{2\pi} d\psi \left[
    \sum_{m=-\infty}^{\infty} m \, e^{-jm (\psi^{\prime} - \psi)}
  \right] \tilde{\mathcal{F}}(\vecbf{r},\vecbf{R}) \, .
\end{equation}
The quantity in the square brackets is
\begin{equation} \label{Eq:DD}
\sum_{m=-\infty}^{\infty} m \, e^{-jm (\psi^{\prime} - \psi)} \, = \,
  j \, 2\pi \, \delta^{\prime}(\psi^{\prime} - \psi) \, ,
\end{equation}
which is proportional to the derivative of another delta function identity.
Since
\begin{equation} \label{Eq:DDInt}
\int dx \, g(x) \, \delta^{\prime}(x-x_{0}) \, = \,
  \left. - g^{\prime}(x) \right|_{x=x_{0}} \, ,
\end{equation}
Equation \ref{Eq:BTildeMMTotal2Summed} then becomes
\begin{eqnarray} \label{Eq:Torque4}
\tilde{l}_{Z} &=
&-j \int d^{2}R
  \left.
    \frac{
      \partial \tilde{\mathcal{F}}(\vecbf{r},\vecbf{R})
    }{
      \partial \psi^{\prime}
    }
  \right|_{\psi^{\prime} = \psi} / \tilde{B} \, = \,
-j \int d^{2}R \,
  \left.
    \frac{\partial s(\vecbf{r})}{\partial \psi^{\prime}}
  \right|_{\psi^{\prime} = \psi} s^{*}(\vecbf{R}) / S \, + \,
  \int d^{2}\Omega \int d^{2}R \, \left[2\pi R \rho \sin{(\phi-\psi)}\right] \,
  B(\vecbf{\Omega}) \, S(\vecbf{R}) / \tilde{B} \nonumber \\ &=
&\sum_{m=-\infty}^{\infty} m \, p^{a}_{m,m} \, + \,
  2\pi \int d^{2}\Omega \int d^{2}R \,
  \left[ \vecbf{\Omega} \times \vecbf{R} \right] \,
  p(\vecbf{\Omega}) \, p^{a}(\vecbf{R}) \, ,
\end{eqnarray}
where the required variables and functions are defined in Appendix
\ref{App:IllForm}.  This equation is identical to the third illustrative form.
The third illustrative form was derived from the other two in Appendix
\ref{App:IllForm}, so there is no need to rederive them here.  \textbf{Q.E.D.}
\end{mathletters}

\section{Interpolation with No POAM Modulation} \label{App:Interpolate}

In this appendix, I assume aperture function interpolation of the form
\begin{equation} \label{Eq:InterpForm}
s(\vecbf{R}) \, \rightarrow \, \tilde{s}(\vecbf{R}) \, = \,
  \int d^{2}\mathcal{R} \, \mathcal{K}(\vecbf{R},\vecbf{\mathcal{R}}) \,
  s(\vecbf{\mathcal{R}}) \, ,
\end{equation}
where $\mathcal{K}(\vecbf{R},\vecbf{\mathcal{R}})$ is the interpolation kernel,
$\vecbf{R}$ $=$ $(R\cos{\psi},R\sin{\psi})$ is the interpolated aperture
coordinate, and $\vecbf{\mathcal{R}}$ $=$
$(\mathcal{R}\cos{\Psi},\mathcal{R},\sin{\Psi})$ is the real aperture
coordinate.  Both coordinates are in units of wavelength.  This equation can be
expanded into POAM components
\begin{mathletters}
\begin{equation} \label{Eq:InterpPOAM}
\tilde{s}(\vecbf{R}) \, = \,
  \sum_{n=-\infty}^{\infty} \tilde{s}_{n}(R) \, e^{jn\psi}
~~ \stackrel{\mathcal{F}}{\Leftrightarrow} ~~
\tilde{s}_{n}(R) \, = \,
\frac{1}{2\pi} \int_{0}^{2\pi} d\psi \, e^{-jn\psi} \, \tilde{s}(\vecbf{R})
  \, = \,
\sum_{m=-\infty}^{\infty} 2\pi \int_{0}^{R_{limit}} d\mathcal{R} \,
  \mathcal{R} \, \mathcal{K}^{-m}_{n}(R,\mathcal{R}) \, s_{m}(\mathcal{R}) \, ,
\end{equation}
where $R_{limit}$ is the radial integration limit ($R_{tel}$ for single
telescopes or $R_{int}$ for interferometers),
\begin{equation} \label{Eq:KernelPOAM}
\mathcal{K}(\vecbf{R},\vecbf{\mathcal{R}}) \, = \,
  \sum_{n=-\infty}^{\infty} \sum_{m=-\infty}^{\infty}
  \mathcal{K}^{-m}_{n}(R,\mathcal{R}) \, e^{-jm\Psi} \, e^{jn\psi}
~~ \stackrel{\mathcal{F}\mathcal{F}}{\Leftrightarrow} ~~
\mathcal{K}^{-m}_{n}(R,\mathcal{R}) \, = \,
  \frac{1}{2\pi} \int_{0}^{2\pi} d\psi \, e^{-jn\psi} \,
  \frac{1}{2\pi} \int_{0}^{2\pi} d\Psi \, e^{jm\Psi} \,
  \mathcal{K}(\vecbf{R},\vecbf{\mathcal{R}})
\end{equation}
is the double Fourier series expansion of the interpolation kernel, and
\begin{equation} \label{Eq:InterpPOAM2}
s(\vecbf{\mathcal{R}}) \, = \,
  \sum_{m=-\infty}^{\infty} s_{m}(\mathcal{R}) \, e^{jm\Psi}
~~ \stackrel{\mathcal{F}}{\Leftrightarrow} ~~
s_{m}(\mathcal{R}) \, = \,
\frac{1}{2\pi} \int_{0}^{2\pi} d\Psi \, e^{-jm\Psi} \, s(\vecbf{\mathcal{R}})
\end{equation}
is the POAM expansion of the non-interpolated aperture function.  All of these
equations have been derived using POAM calculi \citep{Elias08}.
\end{mathletters}

By definition, a system that doesn't modulate the POAM spectrum means that a
single input POAM state gives rise only to the same output POAM state
\citep{Elias08}, so the interpolation kernel POAM gain must be
$\mathcal{K}^{-m}_{n}(R,\mathcal{R})$ $=$ $K_{m}(R,\mathcal{R})$ $\delta_{n,m}$.
The interpolation kernels that produce such POAM gains consist of the family of
functions that exhibit circular symmetry, or
\begin{equation} \label{Eq:KKK}
\mathcal{K}(\vecbf{R},\vecbf{\mathcal{R}}) \, = \,
\mathcal{K}(\left|\vecbf{R}-\vecbf{\mathcal{R}}\right|) \, = \,
\mathcal{K}(\sqrt{R^{2}+\mathcal{R}^{2}-2R\mathcal{R}\cos{(\psi-\Psi)}}) \, .
\end{equation}
To prove this assertion, I substitute this kernel into Equation
\ref{Eq:KernelPOAM} and obtain
\begin{mathletters}
\begin{equation} \label{Eq:KernelPOAMCirc}
\mathcal{K}^{-m}_{n}(R,\mathcal{R}) \, = \,
\left[
  \frac{1}{2\pi} \int_{0}^{2\pi} d\chi \, e^{-jm\chi} \,
  \mathcal{K}(\sqrt{R^{2}+\mathcal{R}^{2}-2R\mathcal{R}\cos{(\psi-\Psi)}})
\right] \,
\left[ \frac{1}{2\pi} \int_{0}^{2\pi} d\psi \, e^{-j(n-m)\psi} \right] \, = \,
  K_{m}(R,\mathcal{R}) \, \delta_{n,m} \, ,
\end{equation}
which means that
\begin{equation} \label{Eq:SnTildeSn}
\tilde{s}_{n}(R) \, = \,
  2\pi \int_{0}^{R_{instr}} d\mathcal{R} \, \mathcal{R} \,
  K_{n}(R,\mathcal{R}) \, s_{n}(\mathcal{R}) \, .
\end{equation}
I employed a change of integration variable $\Psi$ $=$ $\psi$ $-$ $\chi$ and
eliminated $\psi$ from both $\chi$ integration limits.
\textbf{Q.E.D.}
\end{mathletters}

This type of circularly symmetric interpolation kernel is often used in
interferometric imaging, which is indeed fortunate.  According to Equation
\ref{Eq:SnTildeSn}, it appears that the $s_{n}(\mathcal{R})$ must be calculated
in order to determine the $\tilde{s}_{n}(R)$.  This procedure is unnecessarily
complicated.  If one substitutes Equation \ref{Eq:InterpPOAM2} into Equation
\ref{Eq:SnTildeSn}, one finds that
\begin{equation} \label{Eq:SnTildeS}
\tilde{s}_{n}(R) \, = \,
  \int d^{2}\mathcal{R} \, \left[K_{n}(R,\mathcal{R}) \, e^{-jn\Psi} \right] \,
  s(\vecbf{\mathcal{R}}) \, .
\end{equation}
In other words, the $n^{th}$ interpolated POAM gain can be calculated directly
from the continuous or discrete aperture function and the kernel contained in
the square brackets.

\section{Telescope Apodization in Interferometers} \label{App:ApodTel}

The interferometric PSF in Equation \ref{Eq:TelResponseA} can be expanded as a
interferometric DF dual azimuthal Fourier series squared, or
\begin{equation} \label{Eq:PIntExpApod}
P(\vecbf{\Omega}^{\prime},\vecbf{\Omega}) \, = \,
\left|D(\vecbf{\Omega}^{\prime},\vecbf{\Omega})\right|^{2} \, = \,
\left|
  \sum_{p=-\infty}^{\infty} \sum_{k=-\infty}^{\infty}
  D_{p}^{-k}(\rho^{\prime},\rho) \, e^{-jk \phi} \, e^{jp \phi^{\prime}}
\right|^{2} \, = \,
\sum_{p=-\infty}^{\infty} \sum_{q=-\infty}^{\infty}
  \sum_{k=-\infty}^{\infty} \sum_{l=-\infty}^{\infty}
  D_{p}^{-k}(\rho^{\prime},\rho) \, D_{q}^{-l *}(\rho^{\prime},\rho) \,
  e^{-j(k-l)\phi} \, e^{j(p-q)\phi^{\prime}} \, ,
\end{equation}
where $D_{m}^{-n}(\rho^{\prime},\rho)$ is defined in Equation \ref{Eq:IOgainDF}.
Similarly, the source intensity can be expanded into its POAM correlation
expansion
\begin{equation} \label{Eq:BSourceExpApod}
B(\vecbf{\Omega}) \, = \,
  \sum_{m=-\infty}^{\infty} \sum_{n=-\infty}^{\infty}
  B_{m,n}(\rho) \, e^{j(m-n)\phi} \, .
\end{equation}
When these equations are substituted into Equation \ref{Eq:TelResponseA}, the
indices rearranged, and the integral over $\phi$ performed, Equations
\ref{Ml:ApodResponsePOAM}b-c result.
\textbf{Q.E.D.}


\begin{thebibliography}{}
\bibitem[Elias(2008)]{Elias08} Elias II, N.M. 2008, \aap, 492, 883.
\bibitem[Harwit(2003)]{Harwit03} Harwit, M. 2003, \apj, 597, 1266.
\bibitem[Brentjens(2011)]{Brentjens} Brentjens, M. 2011, Unpublished EVLA
  holography data.
\bibitem[Rau \etal(2009)]{UrvashiIEEE} Rau, U., Bhatnagar, S., Voronkov, M.A.,
  Cornwell, T.J. 2009, IEEE, 97, 1472.
\bibitem[Rau(2010)]{UrvashiPhD} Rau, U. 2010, PhD Thesis, New Mexico Institute
of Mining and Technology.
\bibitem[Tamburini \etal(2006)]{Tamburini} Tamburini, F., Anzolin, G.,
  Umbriaco, G., Bianchini, A., and Barbieri, C. 2006, Phys.Rev.Lett., 97,
  163903.
\bibitem[Tamburini \etal(2011)]{TTM-TA} Tamburini, F., Thid{\'e}, B.,
  Molina-Terriza, G., Anzolin, G. 2011, Nat.Phys., 7, Issue 3, 195.
\end{thebibliography}
\end{document}